\documentclass[aps,twocolumn,prd]{revtex4-1}

\usepackage{graphicx}
\graphicspath{{../plots/}}
\usepackage{appendix}
\usepackage{natbib}
\usepackage[breaklinks,colorlinks,citecolor=blue]{hyperref}
\usepackage{amsmath}
\usepackage{mathtools}

\input epsf

\def\physrep{Physics Reports }
\def\jcap{JCAP }

\def\be{\begin{equation}}
\def\ee{\end{equation}}
\def\bea{\begin{align}}
\def\eea{\end{align}}

\def\cmm2{{\,\rm cm^{-2}}}
\def\cm2{{\,{\rm cm}^2}}
\def\cmm3{{\,{\rm cm}^{-3}}}
\def\gcmm3{{\,{\rm g\,cm^{-3}}}}

\def\fun#1#2{\lower3.6pt\vbox{\baselineskip0pt\lineskip.9pt
  \ialign{$\mathsurround=0pt#1\hfil##\hfil$\crcr#2\crcr\sim\crcr}}}

\def\c{{\cal C}}

\def\p3m{P$^3$M}

\def\ga{\mathrel{\mathpalette\fun >}}
\def\fun#1#2{\lower3.6pt\vbox{\baselineskip0pt\lineskip.9pt
  \ialign{$\mathsurround=0pt#1\hfil##\hfil$\crcr#2\crcr\sim\crcr}}}

\def\aap{Astron. \& Astrophys.}

\begin{document}

\def\affilucd{\affiliation{Department of Physics, One Shields Avenue,
University of California, Davis, California 95616, USA}}
\def\affiluofi{\affiliation{Departments of Astronomy and of Physics,
University of Illinois,Urbana-Champaign, IL, USA}}

\title{New Bounds for Axions and Axion-Like Particles with keV-GeV Masses}
\author{Marius\ Millea}
\affilucd
\author{Brian\ Fields}
\affiluofi
\author{Lloyd\ Knox}
\affilucd

\begin{abstract}
We give updated constraints on hypothetical light bosons with a two-photon coupling such as axions or axion-like particles (ALPs). We focus on masses and lifetimes where decays happen near big bang nucleosynthesis (BBN), thus altering the baryon-to-photon ratio and number of relativistic degrees of freedom between the BBN epoch and the cosmic microwave background (CMB) last scattering epoch, in particular such that $N_{\rm eff}^{\rm CMB} < N_{\rm eff}^{\rm BBN}$ and $\eta^{\rm CMB} < \eta^{\rm BBN}$. New constraints presented here come from {\it Planck} measurements of the CMB power spectrum combined with the latest inferences of primordial $^4$He and D/H abundances. We find that a previously allowed region in parameter space near $m=1\,\rm MeV$ and $\tau=100\,\rm ms$, consistent with a QCD axion arising from a symmetry breaking near the electroweak scale, is now ruled out at $>3\sigma$ by the combination of CMB+D/H measurements if only ALPs and three thermalized neutrino species contribute to $N_{\rm eff}$. The bound relaxes if there are additional light degrees of freedom present which, in this scenario, have their contribution limited to $\Delta N_{\rm eff}=1.1\pm0.3$. We give forecasts showing that a number of experiments are expected to reach the sensitivity needed to further test this region, such as Stage-IV CMB and SUPER-KEKB, the latter a direct test insensitive to any extra degrees of freedom. 
\end{abstract}
  \maketitle


\section{Introduction}  

Recent improvements in both measurements of the Cosmic Microwave Background (CMB) angular power spectrum and inferences of primordial elemental abundances formed during Big Bang Nucleosynthesis (BBN) motivate us to reconsider bounds on hypothetical scenarios which alter the expansion rate or inject energy into the plasma around these two epochs. One of the simplest scenarios involves a radiatively decaying particle, for which axions or so called ``axion-like particles'' (ALPs) provide a theoretically well motivated candidate. Axions arise from perhaps the most elegant solution to the strong CP problem as the pseudo Nambu-Goldstone boson of a new spontaneously broken symmetry \citep{peccei1977,peccei1977a,wilczek1978,weinberg1978}. Axions have a mass and standard model couplings controlled by a single parameter: the energy scale of the symmetry breaking. ALPs form a more general class of particles where the mass and couplings are independent. Such models can arise from other new symmetries which are spontaneously broken \citep{chikashige1981,gelmini1981}, and in string theory \citep{arvanitaki2009}.

In the parameter space of interest here, ALPs are weakly interacting, making them difficult to detect in the laboratory. Cosmological constraints serve as a natural complement as the weak coupling generally leads to later decays allowing the particles to become non-relativistic and pick up energy compared to the plasma, leading to observable consequences upon their decay. Astrophysical bounds are also important as ALPs provide a new method for energy release from stars and supernovae. Some early calculations and compilation of cosmological, astrophysical, and laboratory bounds on ALPs include those from \citet{masso1995,masso1997}. Cosmological bounds were recently updated by \citet{cadamuro2011a} which considered only axions, and \citet{cadamuro2012} who extended this more generally to ALPs. Among other significant advances, \citet{cadamuro2012} used newer data, treated out-of-equilibrium decays more carefully, and performed precise calculations of the implications for BBN. These and other known ALP bounds are tabulated in \cite{hewett2012,essig2013,olive2014}. Our work updates these by 1) using the latest inferences of primordial element abundances, 2) using the latest measurements of the CMB power spectrum measurements from {\it Planck}, 3) having improved calculations of CMB spectral distortions.

We also highlight the importance of a region in parameter space near $m=1\,\rm MeV$ and $\tau=100\,\rm ms$ which we term the MeV-ALP window. This region is interesting because it previously has evaded all known constraints (also noted in \cite{hewett2012,mimasu2014}) and, as we show, can correspond to a particular axion model we will call the DFSZ-EN2. Additionally, in this mass window the symmetry breaking scale is much lower than the often-considered ``invisible'' axion models. A main conclusion of this paper is that the possibility of the DFSZ-EN2 or any other ALP hiding in plain sight in the MeV-ALP window is ruled out by the newer bounds presented here. We show, however, that these bounds are model dependent and the MeV-ALP window can be reopened if there is other exotic radiation present. 

The paper is structured as follows. In Sec.~\ref{sec:scenario} we discuss in more detail the scenario and its implications for cosmology. Sec.~\ref{sec:constraints} describes the new and tabulated constraints which we use.  In Sec.~\ref{sec:axion} we further discuss the MeV-ALP window and in Sec.~\ref{sec:forecasts} we give forecasts for future probes.

\section{The Scenario}
\label{sec:scenario}
We begin by discussing the cosmological impact of ALPs, which we define as any particle with a mass $m_\phi$ and two-photo coupling $g_{\phi\gamma}$. Following standard conventions in the ALP literature, the effective Lagrangian is 
\begin{align}
\label{eq:lagrangian}
\mathcal{L} = \frac{1}{2}(\partial_\mu \phi)(\partial^\mu \phi) - \frac{1}{2}m_\phi^2\phi^2 - \frac{g_{\phi\gamma}}{4}\phi F_{\mu\nu}\tilde{F}^{\mu\nu},
\end{align}
where $F$ is the electromagnetic field strength tensor, $\tilde F$ its dual, and $\phi$ the ALP field. We often describe the two dimensional parameter space with $m_\phi$ and, in place of $g_{\phi \gamma}$, the lifetime for decay into photons
\begin{align}
\tau_{\phi\gamma} \equiv \Gamma^{-1}_{\phi\gamma} = \frac{64\pi}{m_\phi^3g_{\phi\gamma}^2}.
\end{align}

Two processes drive the cosmological evolution of the ALP energy density. The first is the Primakoff interaction which allows for conversion between photons and ALPs in the presence of a charged particle $q$, via $\gamma q\leftrightarrow  \phi q$. Because it is a four-point interaction, the scattering rate for the Primakoff process will depend on the density of scatterers (in this case charged particles). At early times, this density grows faster than the Hubble rate, meaning the Primakoff process will always begin in equilibrium and freeze out at later times. The second process is the direct two-photon interaction, $\gamma\gamma\leftrightarrow \phi$. Conversely, this three-point interaction has no dependence on scatterers, and will always begin out of equilibrium then re-equilibrate at later times. 

Qualitatively, the ALP scenario depends strongly on the time ordering in which ALPs 1) freeze-out from the Primakoff interaction, 2) become non-relativistic, and 3) recouple via the two-photon interaction. The details of how these events are controlled by the two free parameters forbids certain orderings, and in fact the vast majority of solutions fall into one of just two cases. If there is a gap between freeze-out and recoupling during which the ALPs become non-relativistic, meaning $T_{\rm fo} > m_\phi > T_{\rm re}$, there is an out-of-equilibrium decay. In this case, upon becoming non-relativistic ALPs cease to track their equilibrium abundance and instead increase in energy density relative to the plasma. Decay happens when the two-photon interaction becomes effective, which here is controlled by only the ALP lifetime, independent of the mass. On the other hand, if ALPs become non-relativistic only after recoupling, $T_{\rm re}>m_\phi$, they will track their equilibrium abundance throughout decay, a scenario which might better be called a ``Boltzmann suppression.'' This suppression occurs when the temperature reaches the mass, independent of the lifetime. Although some subtleties can occur if other events reheat the plasma while the ALPs are decoupled, to a large degree only these two in- and out-of-equilibrium scenarios are important.

In discussing the cosmological impact of ALPs, it is useful to define two quantities. The first is the effective number of relativistic species, $N_{\rm eff}$. As usual, this is taken so that in the limit of complete neutrino decoupling prior to electron-positron annihilation, each neutrino species (with antineutrinos) contributes 1 to $N_{\rm eff}$, making the total relativistic energy density, 
\begin{align}
\label{eq:neff}
\rho_{\rm rel} = \rho_\gamma \left[ 1 + N_{\rm eff} \frac{7}{8}\left(\frac{4}{11}\right)^{\frac{4}{3}} \right].
\end{align}
The other quantity is the baryon-to-photon ratio $\eta$, or equivalently the energy density in baryons $\Omega_bh^2$, 
\begin{align}
\eta = \frac{n_b}{n_\gamma} = 2.74 \times 10^{-8} \Omega_bh^2.
\end{align}
We sometimes superscript these quantities with BBN or CMB depending on the epoch at which they are evaluated, although we note there is no exact definition as they can change with time in this scenario. 

ALP decays can only reduce or leave unchanged the value of $N_{\rm eff}^{\rm CMB}$. If the
decays occur while the neutrinos are still fully coupled then $N_{\rm
  eft}^{\rm CMB}$ does not change. Later decays increase the
temperature of the plasma relative to the neutrinos and thus decrease
$N_{\rm eff}^{\rm CMB}$. The value of $N_{\rm eff}^{\rm BBN}$ can also
be decreased in the same manner if decay happens before BBN, although
this region in parameter space is fairly small. A much larger region
of parameter space corresponds to ALPs decaying after BBN and
increasing $N_{\rm eff}^{\rm BBN}$ by simply contributing to the
relativistic energy density during BBN. ALP decays affect $\eta$
similarly to $N_{\rm eff}$ by causing its value to be less after the
decay. The difference is that with $\eta^{\rm CMB}$ held constant,
$\eta$ before the decay is now increased. Assuming that the physics of
BBN is unchanged, both an increase in $\eta_{\rm BBN}$ and an increase in
$N_{\rm eff}^{\rm BBN}$ serve to increase the amount of primordial
helium produced. For deuterium, however, there is partial
cancellation, leaving the primordial abundance only slightly
reduced. This cancellation is key to allowing the ALP scenario at all
in light of the very tight constraints on primordial deuterium (see
Sec.~\ref{sec:bbnconstraints}). The increase in primordial helium
coupled with the decrease in $N_{\rm eff}^{\rm CMB}$ moves along a
degeneracy direction for the CMB constraints, and is hence also
generally allowed.

More quantitatively, we must track the evolution of the phase space distribution function $f_\phi$, which is governed by the Boltzmann equation
\begin{align}
\frac{df_\phi}{dt}= (C_q + C_\gamma) (f_\phi^{\rm eq} - f_\phi),
\label{eq:boltz}
\end{align}
where the $f$'s are comoving, $f_\phi^{\rm eq}$ is a Bose-Einstein distribution, and a dependence on the comoving momentum $p$ is omitted for brevity. When electrons and positrons are the only charged particles present, the scattering rate for the Primakoff interaction, $C_q$, is given by
\begin{align}
\label{eq:primakoff}
C_q &\approx n_{ep} g_{\phi\gamma}^2 \frac{\alpha}{16} \log\left\{1+\frac{\left[4E(m_e + 3T)\right]^2}{m_\gamma^2\left[m_e^2 + (m_e+3T)^2\right]} \right\},
\end{align}
where $n_{ep}$ is the number density of electrons and positrons, $E$ is the ALP energy, and $m_\gamma=eT/3$ is the plasmon mass in an electron-positron plasma \citep{bolz2001a}. As shown by \citet{cadamuro2012}, an accounting of other charged particles gives an approximation for the Primakoff freeze-out temperature of
\begin{align}
\label{eq:freezeout}
T_{\rm fo} \approx 123 \frac{\sqrt{g_*(T_{\rm fo})}}{g_q(T_{\rm fo})} \left( \frac{10^{-9} {\rm GeV}^{-1}}{g_{\phi\gamma}} \right)^2 \,\rm GeV,
\end{align}
where $g_*$ is the number of relativistic degrees of freedom in the plasma, and $g_q$ the number of charged relativistic degrees of freedom \footnote{The degrees of freedom $g_q$ are counted identically to $g_*$ but weighted by the square of the charge of each species, in units of the elementary charge. Eqn.~\ref{eq:primakoff} ignores the dependence of the plasmon mass on $g_q$, leading to changes in freeze-out temperature as large as $\sim 25\%$ \cite[see][]{cadamuro2012}. Since the details of Primakoff freeze-out are not important to our main conclusions, this is sufficient.}. 

\begin{figure}
\includegraphics[width=3in]{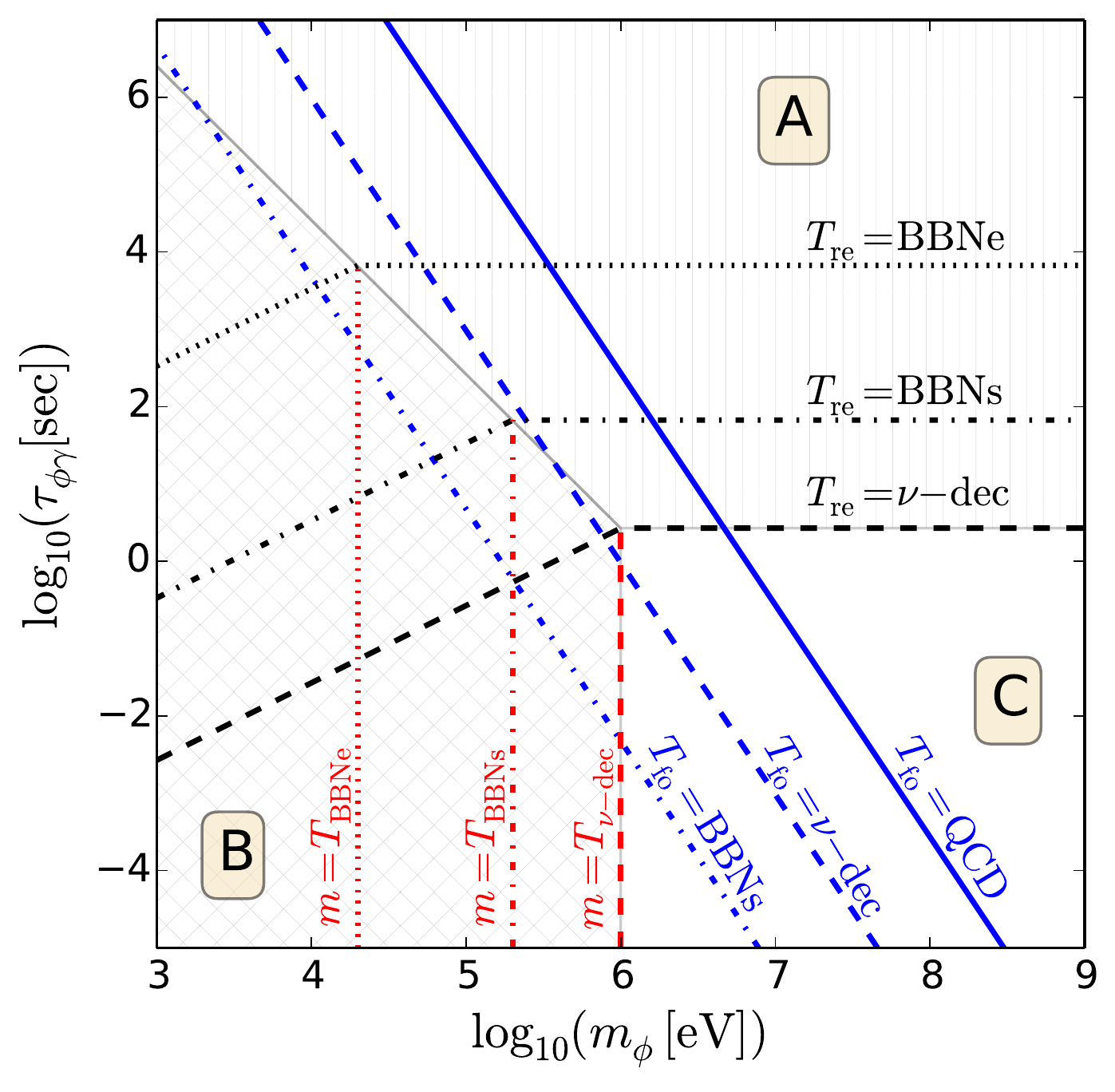}
\caption{Key regions and contours in the mass-lifetime parameter space according to the analytic approximations in Sec.~\ref{sec:scenario}. As in other plots in this paper, dashed lines correspond to the temperature at neutrino decoupling, dot-dashed lines the start of BBN, and dotted lines the end of BBN. Blue lines show contours of constant Primakoff freeze-out temperature, $T_{\rm fo}$, and black lines show contours of constant two-photon re-equilibration temperature, $T_{\rm re}$. The line $T_{\rm re}=m_\phi$ divides two regions A and B. Region A is vertically hatched and corresponds to out-of-equilibrium decays. Region B is cross hatched and corresponds to in-equilibrium decays. Constant decay-time contours in region A are $T_{\rm re}=\rm const$ whereas they are $m_\phi=\rm const$ in region B. Region C has no hatching and corresponds to decays before neutrino decoupling, where ALPs leave no cosmologically observable traces. The line $T_{\rm fo}$=QCD leaves a sharp feature on cosmological constraints as $g_*$ changes suddenly during this phase transition.}
\label{fig:mtau_regions}
\end{figure}

The two-photon scattering rate, $C_\gamma$, can be calculated exactly at all times,
\begin{align}
\label{eq:twophoton}
C_\gamma &= \frac{m_\phi}{E \tau_{\phi\gamma}} \left[ \frac{2T}{p} \ln \frac{\sinh \frac{E+p}{4T}}{\sinh \frac{E-p}{4T}} \right].
\end{align}
Here we have ignored the plasmon mass, which inhibits decays when $m_\phi<2m_\gamma$, as this effect is not important in the parameter range of interest. We have also assumed that photon rethermalization is instantaneous, which is an excellent approximation at early times. We consider thermalization at later times in Sec.~\ref{sec:cmbspec}. 

In practice, we begin evolving the Boltzmann equations at a temperature $T_0$ when electrons and positrons are the only remaining charged particles and we can thus use Eqn.~\ref{eq:primakoff} for the Primakoff scattering rate. If the Primakoff process has yet to freeze out at $T_0$, as given by Eqn.~\ref{eq:freezeout}, we take take as the initial conditions for $f_\phi$ a Bose-Einstein distribution at temperature $T_0$. If freeze-out is earlier, we use the conservation of comoving entropy to calculate the increase in photon temperature after Primamkoff freeze-out. The result is that ALPs are now at a reduced relative temperature of $T_0 (g_*(T_0)/g_*(T_{\rm fo}))^{1/3}$.

We now describe key regions in the ALP parameter space using Fig.~\ref{fig:mtau_regions} as a guide. An important quantity governing the ALP evolution is the temperature, $T_{\rm re}$, at which the two-photon interaction re-equilibrates. Eqn.~\ref{eq:boltz} shows that the ALP distribution $f_\phi$ at a given momentum $p$ becomes equal to its equilibrium value $f_\phi^{\rm eq}$ when the scattering rate, $C_\gamma(p)$, is on the order of the Hubble constant. We consider only those momenta which contribute dominantly to the total energy density, since this is the quantity we are interested in. For $m_\phi\gtrsim T$, these are $p\approx 0$. In this case, $C_\gamma$ vastly simplifies to $1/\tau_{\phi\gamma}$ and thus recoupling occurs when the Hubble time reaches the lifetime. For $m_\phi\lesssim T$, the important momenta are instead $E\approx p\approx T$. Ignoring the term in brackets in Eqn.~\ref{eq:twophoton} which can be shown to depend only logarithmically on temperature for these momenta, and using $H=1.66 \sqrt{g_*}T^2$, we arrive at (in Planck units),
\begin{align}
\label{eqn:Tre}
T_{\rm re} =
\begin{dcases}
    \left(\frac{1}{\tau_{\phi\gamma}1.66\sqrt{g_*}}\right)^{1/2} & m_\phi>T_{\rm re} \\
    \left(\frac{m_\phi}{\tau_{\phi\gamma}}\frac{1}{1.66\sqrt{g_*}}\right)^{1/3} & m_\phi<T_{\rm re}
\end{dcases}
\end{align}
Fig.~\ref{fig:mtau_regions} shows constant $T_{\rm re}$ contours in black for three values of $T_{\rm re}$ corresponding to neutrino decoupling, and the start and end of BBN. Here and throughout this paper, we take these temperatures to be 1\,MeV, 200\,KeV, and 20\,KeV respectively. Fig.~\ref{fig:mtau_regions} also shows constant mass contours at these three values.  The line implicitly formed by $T_{\rm re}=m_\phi$ divides regions A and B where the ALP becomes non-relativistic before and after the two-photon interaction re-equilibrates, respectively. Note that in region A we have $m_\phi>T_{\rm re}$, which is only part of the requirement for an out-of-equilibrium decay. The other is that $T_{\rm fo}>m_\phi$, however it turns out that this is always satisfied for any combination of mass and lifetime in region A; that is to say, ALPs can never decay by the Primakoff process alone. 

\begin{figure}
\centering
\includegraphics[width=2.8in]{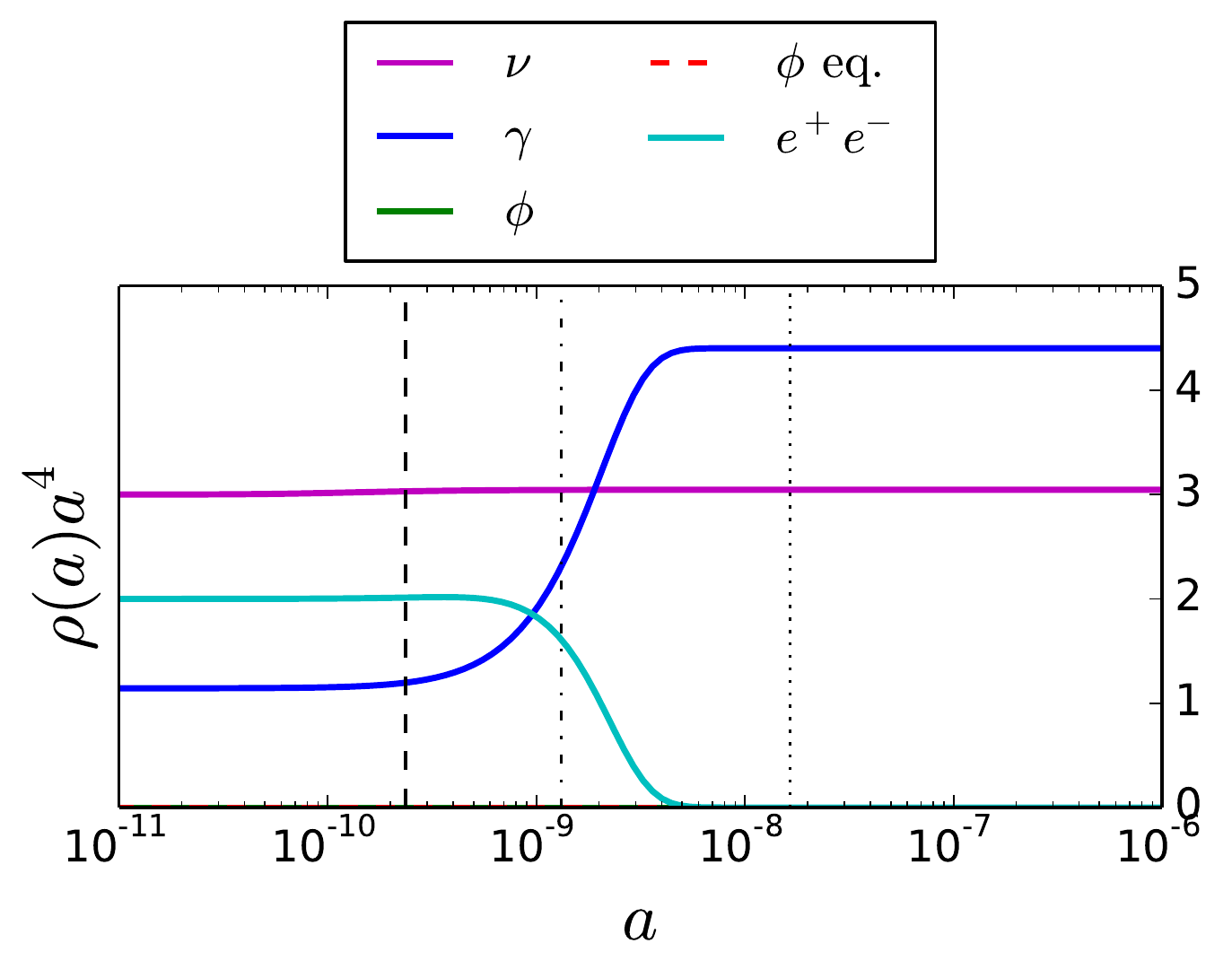}
\includegraphics[width=2.8in]{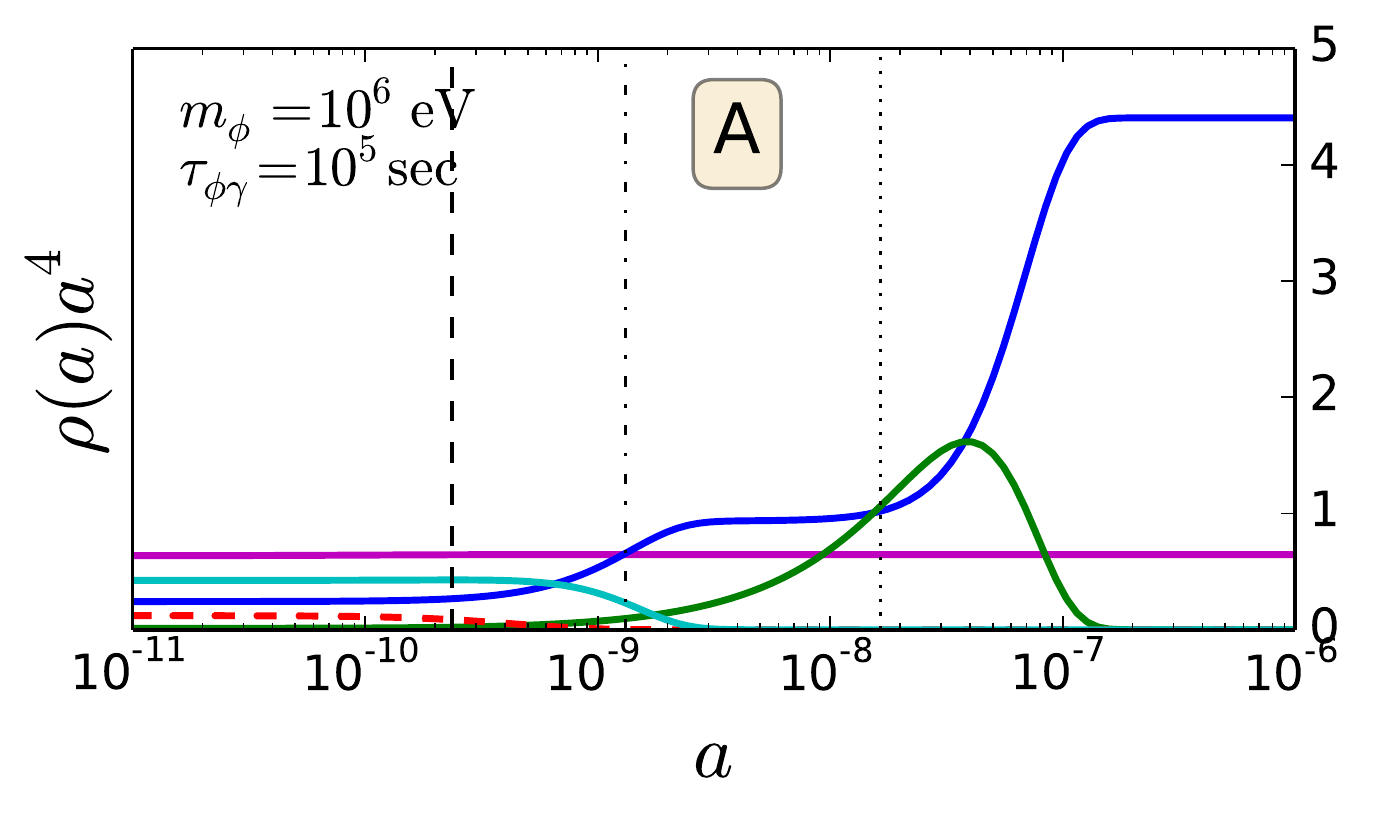}
\includegraphics[width=2.8in]{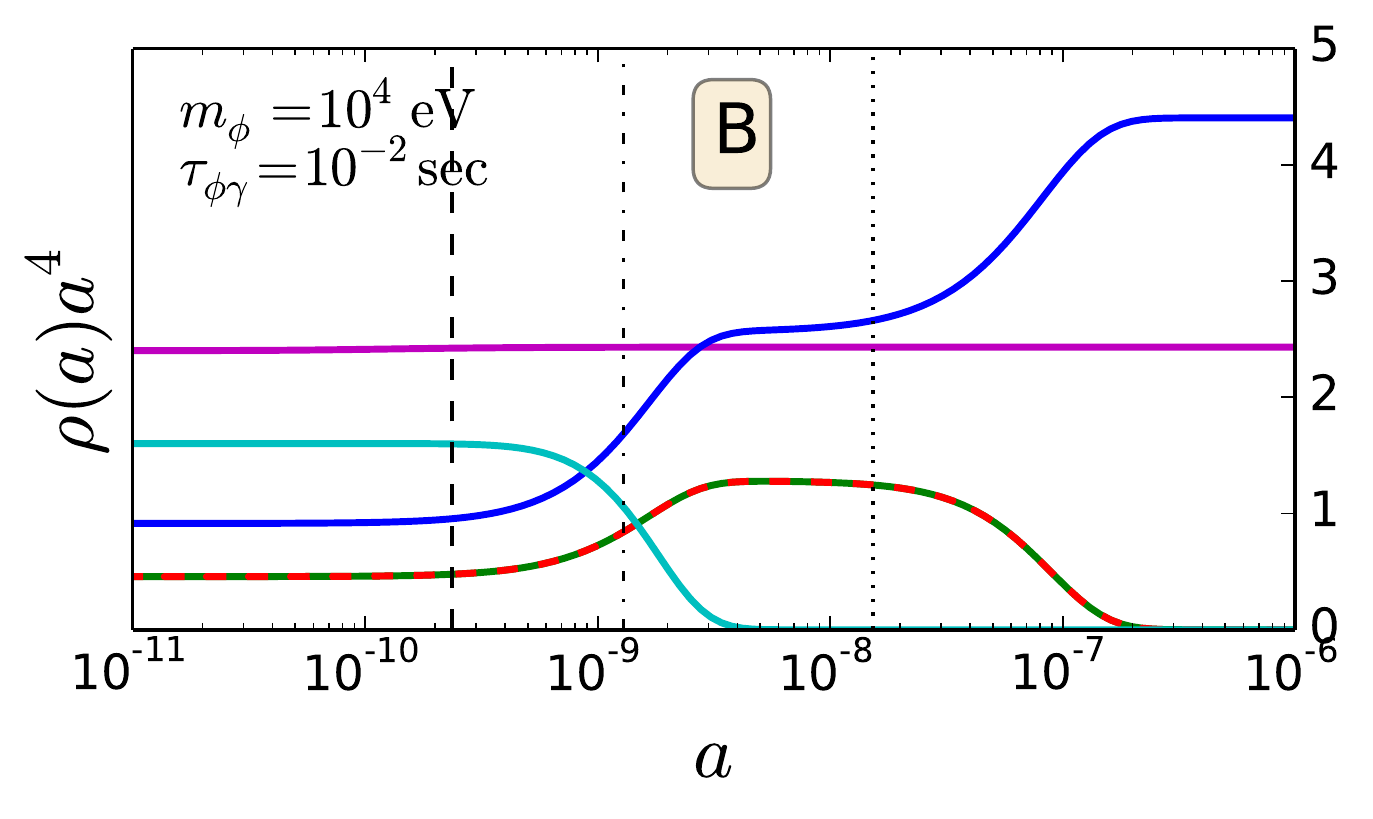}
\includegraphics[width=2.8in]{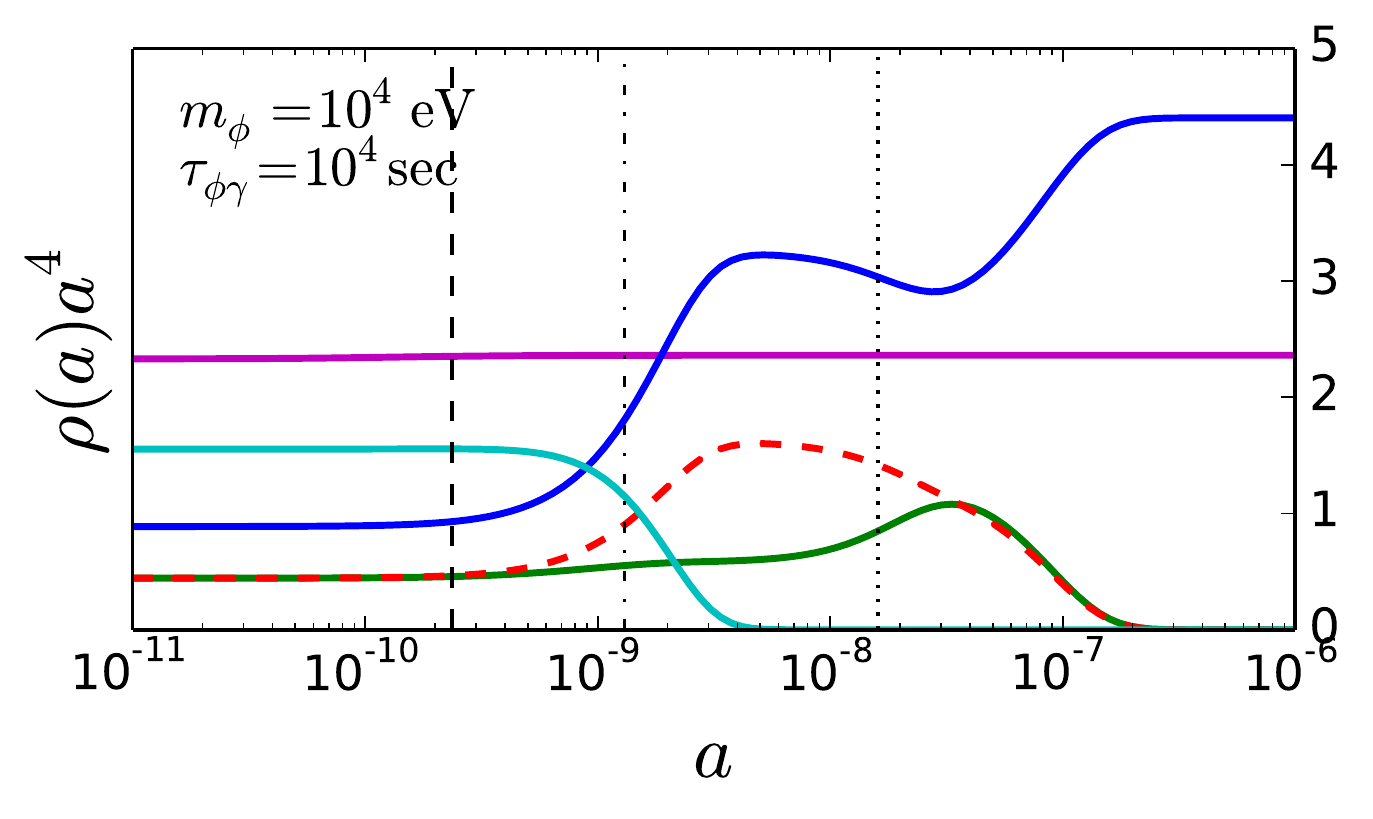}

\caption{The evolution of the energy densities in the various components of the universe for different scenarios which have similar decay time. The temperature of the photons today is held fixed and the y-axis units are such that the final value of the neutrino line is the value of $N_{\rm eff}^{\rm CMB}$. As in other plots in this paper, dashed lines correspond to the scale factor at neutrino decoupling, dot-dashed lines the start of BBN, and dotted lines the end of BBN. The dashed red line is not actually a component, but is shown for illustrative purposes; it is the equilibrium ALP energy density (that is, the energy density APLs would have if they were in chemical and kinetic equilibrium with the photons). As per Eqn.~\ref{eq:boltz}, interactions serve to always drive the ALP energy density towards equilibrium. The plots labeled A and B correspond to the same regions in Fig.~\ref{fig:mtau_regions}.}
\label{fig:evol}
\end{figure}

\begin{figure*}
\includegraphics[width=7in]{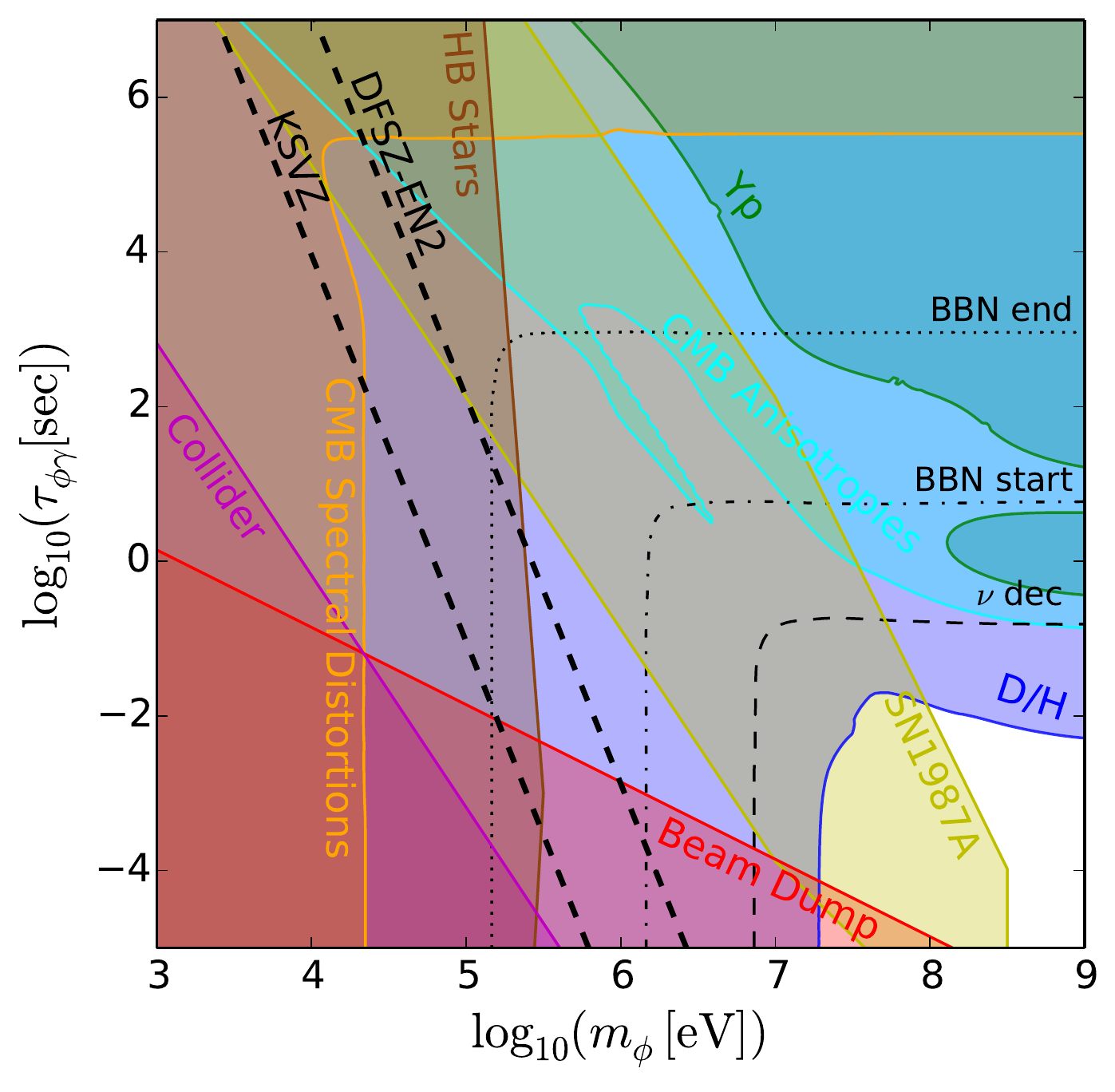}
\caption{Exclusion regions in the ALP mass-lifetime parameter space. The dashed and dotted lines labeled ``$\nu$ dec'' (neutrino decoupling), and ``BBN start/end'' correspond to particles which decay at these particular times (with decay here arbitrarily defined as when maximum energy injection occurs). The two thick dashed lines are the consistency relations for two particular axion models (see Sec.~\ref{sec:axion}). The CMB, D/H, and $Y_p$ regions are excluded at 3$\sigma$, the Collider and Beam Dump regions are excluded at 2$\sigma$, and the SN1987a and HB Stars regions are less formal, rough bounds (see Sec.~\ref{sec:astrophysical}).}
\label{fig:mtau_baseline}
\end{figure*}

\begin{figure*}
\includegraphics[width=3in]{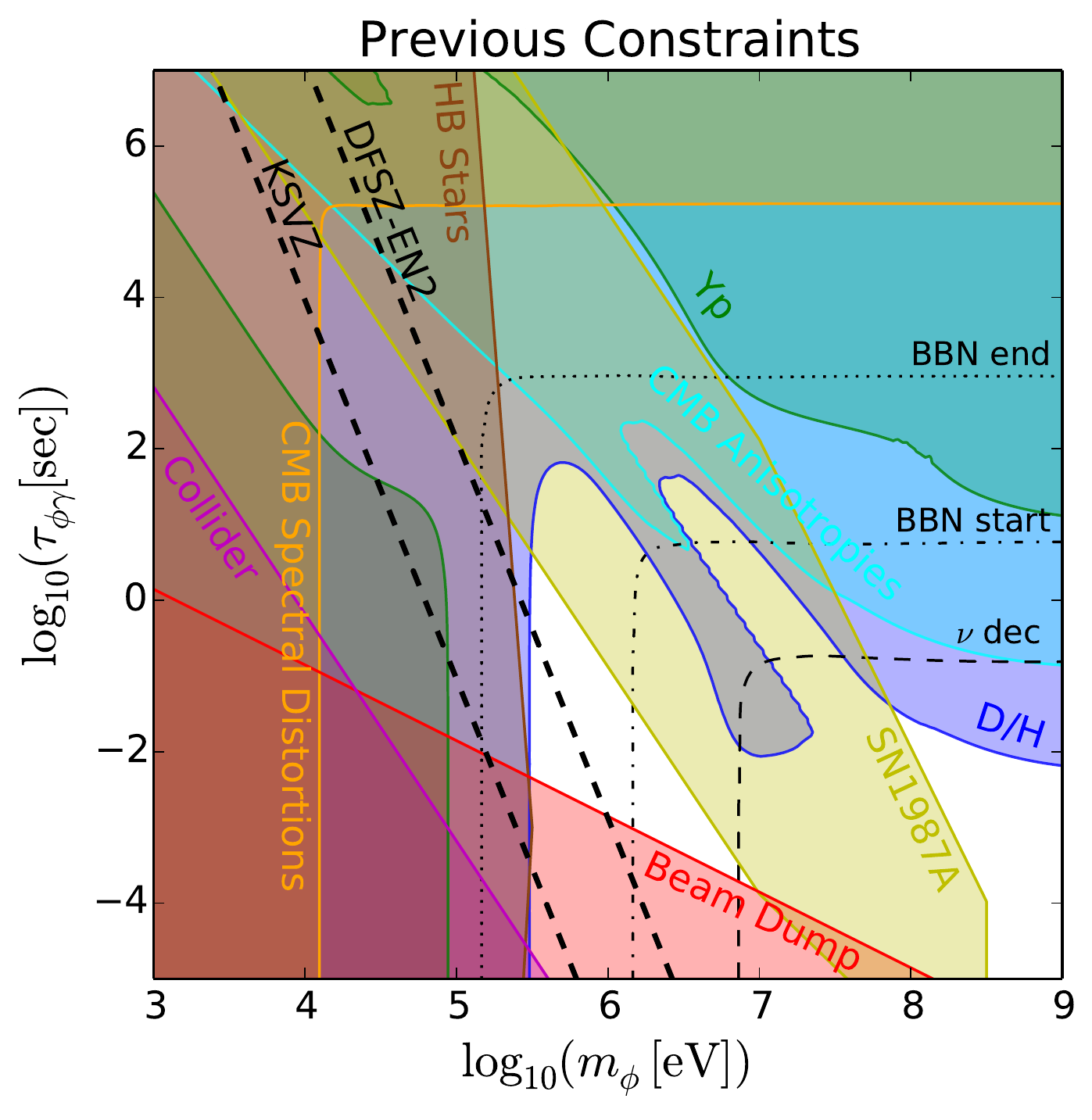}
\includegraphics[width=3in]{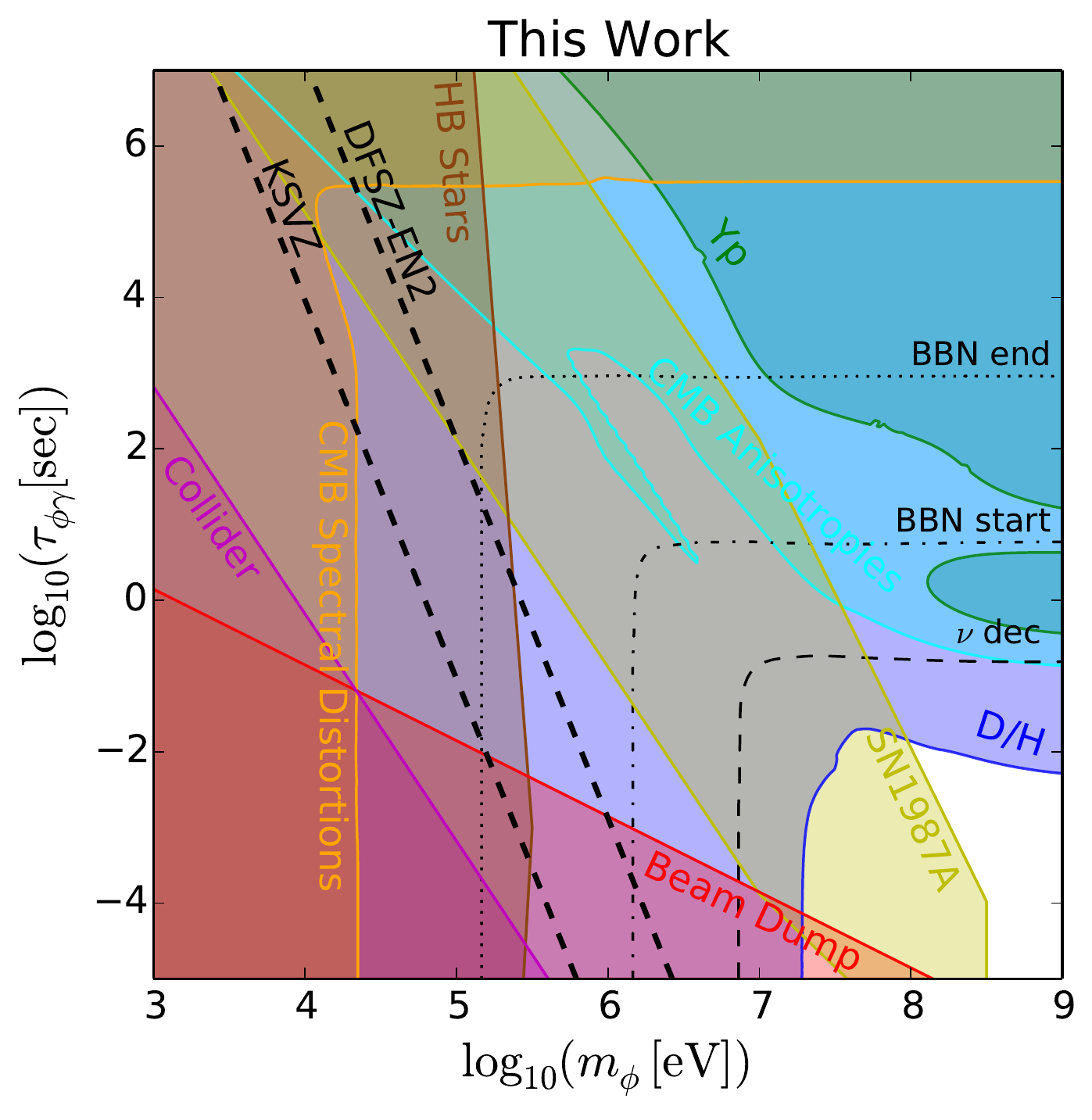}
\caption{A comparison of exclusion regions from previous works (left panel) and those presented here (right panel). The right panel is identical to Fig.~\ref{fig:mtau_baseline}.}
\label{fig:mtau_baseline_compare}
\end{figure*}

\begin{figure*}
\includegraphics[width=7in]{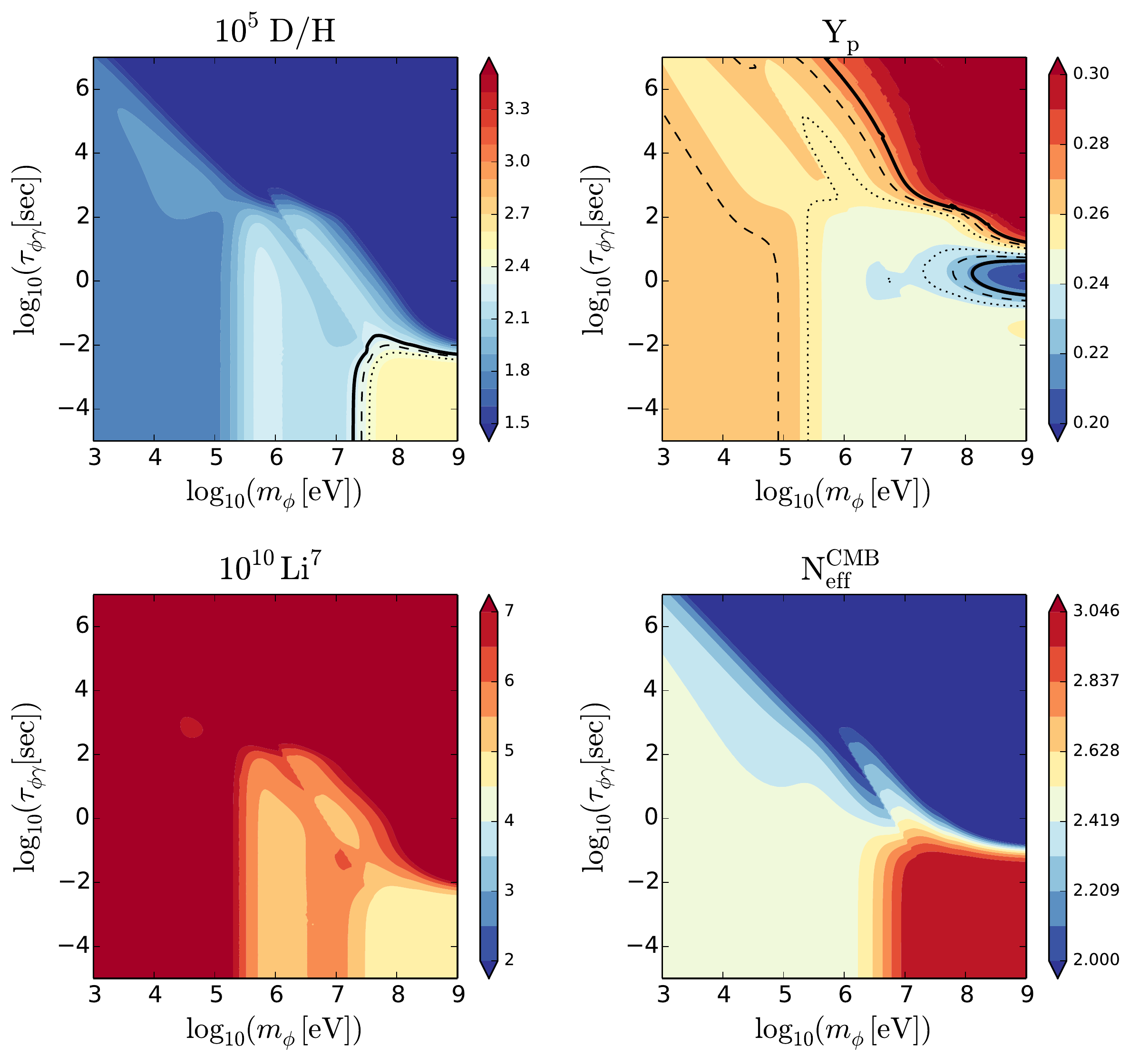}
\caption{The colored contours show the prediction for each of the labeled quantities as a function of different values of ALP mass and lifetime. The dotted/dashed/solid lines give 1/2/3 $\sigma$ contours given the measurements for these quantities discussed in Sec.~\ref{sec:constraints}. No lines are visible on the lithium plot because the entire parameter space is excluded at $>3\sigma$ (our scenario does not alleviate the lithium problem). We do not give contours for the $N_{\rm eff}^{\rm CMB}$ plot because the CMB constraint is highly degenerate with $Y_p$. For the D/H panel, the colored contours are calculated assuming a best-fit $\eta$ from the CMB, and uncertainties in $\eta$ and nuclear reaction rates are taken into account in producing the $\sigma$ contours (see Sec.~\ref{sec:bbnconstraints} for discussion).}
\label{fig:mtau_series_baseline}
\end{figure*}

The evolution of the energy densities in the relevant components of the plasma for a typical in-equilibrium decay (region B) is shown in the third panel of Fig.~\ref{fig:evol}. If ALPs decay after neutrinos are decoupled, conservation of comoving entropy implies that the temperature of the neutrinos relative to the photons after the ALP decay and after electrons and positrons have annihilated is $(4/13)^{1/3}$, as compared to $(4/11)^{1/3}$ in the standard scenario. This means that $N_{\rm eff}^{\rm CMB}$ will be reduced by $(13/11)^{4/3}$ and, at fixed $\eta^{\rm CMB}$, the value of $\eta$ prior to the decay is increased by a factor of $13/11$. Assuming three neutrinos, this gives $N_{\rm eff}^{\rm CMB}\approx2.44$. Because the exact timing with respect to electron-positron annihilation is unimportant for the final temperature, the entirety of region B shares this same value for $N_{\rm eff}^{\rm CMB}$. The BBN data, however, are sensitive to the exact time of decay via sensitivity to $N_{\rm eff}^{\rm BBN}$ and $\eta_{\rm BBN}$. Since decay time when in equilibrium is controlled only by $m_{\phi}$, we find characteristic constant-mass contours in this region in the BBN constraints in Figs.~\ref{fig:mtau_baseline}, \ref{fig:mtau_baseline_compare}, and \ref{fig:mtau_series_baseline}. 

The second panel of Fig.~\ref{fig:evol} shows instead a typical out-of-equilibrium decay (region A). In this region an important quantity is the fractional increase in the energy of the photons (or equivalently the decrease in $N_{\rm eff}$) once the ALP decays at $t\sim\tau_{\phi\gamma}$. Because the ALP here is non-relativistic, this is,
\begin{align}
\label{eq:mroottau}
\frac{1}{N_{\rm eff}}\sim\left.\frac{\rho_\phi}{\rho_\gamma}\right|_{t=\tau_{\phi\gamma}} \sim \left.\frac{m_\phi a^{-3}}{a^{-4}}\right|_{t=\tau_{\phi\gamma}} \sim m_\phi \sqrt{\tau_{\phi\gamma}}
\end{align}
where we have assumed that the ALP does not remain non-relativistic for too long before decay hence the universe is radiation dominated and $a(t)\sim\sqrt{t}$. This assumption is true for models right on the edge of the allowed region, thus we find characteristic contours of constant $m_\phi\sqrt{\tau_{\phi\gamma}}$ for cosmological constraints in region A as seen in Figs.~\ref{fig:mtau_baseline}, \ref{fig:mtau_baseline_compare}, and \ref{fig:mtau_series_baseline}. 

In both regions A and B there are characteristic contours arising from phenomena which depend on the fractional energy injection after a certain reaction has frozen out, i.e. after a certain temperature. For example, CMB and BBN constraints are only sensitive to energy injected after neutrino decoupling, as any earlier injection is ``invisible'' because it is rethermalized among all components. We will also consider bounds from CMB spectral distortions, which depend on the energy injected only after the freeze-out of reactions which can bring the CMB spectrum back into chemical equilibrium. 

In region B, fractional energy injection is independent of either mass or lifetime, and the amount after a certain temperature is controlled only by the mass. On the non-equilibrium side the story is slightly more subtle. The fractional energy increase at a certain time or scale factor depends on the mass, as per Eqn.~\ref{eq:mroottau}. However, we cannot simply assume $a\sim1/T$ because the ALP decay alters this relation. In particular, the Friedmann acceleration equation shows that it does so in a way which exactly cancels the $m_\phi$ dependence, leaving only sensitivity to $\tau_{\phi\gamma}$. Essentially, more massive ALPs lead to more total energy injection, but delay the time it takes to reach a certain temperature, leaving the same amount of energy injected after that temperature. Thus, contours of constant fractional energy injection are $m=\rm const$ in region B and $\tau=\rm const$ in region A. We note that this is the same as a contour of constant decay time. The final key region in Fig.~\ref{fig:mtau_regions}, region C, is delineated by such a contour, and corresponds to zero energy injection after neutrino decoupling and thus no cosmologically observable imprints. 

The Primakoff freeze-out temperature plays a smaller role in the ALP evolution than the recoupling temperature, although we briefly mention two effects stemming from events happening in the gap between freeze-out and recoupling (in regions of parameter space where the gap exists). We first note that if no reheating of the plasma occurs in this gap and ALPs are relativistic, they still track their equilibrium abundance even though they are decoupled. One possible reheating occurring during the gap is the QCD phase transition which imprints a sharp feature our constraints. During this transition the temperature of the ALPs roughly doubles along with the rest of the plasma if they are still coupled, leading to an energy density roughly twenty times larger. Another possibility is reheating from electron-positron annihilation, a case in which the ALPs can re-equilibrate relativistically thereafter. Such a scenario happens near the top-left part of Fig.~\ref{fig:mtau_regions} and is shown in the fourth panel of Fig.\ref{fig:evol}. For deuterium and helium constraints shown in  Fig.~\ref{fig:mtau_series_baseline}, near this region we can find both lines of constant $T_{\rm re}$ and $T_{\rm fo}$.

\section{Constraints}
\label{sec:constraints}

\subsection{Cosmic Microwave Background}
\label{sec:cmb}
\subsubsection{Frequency Spectrum}
\label{sec:cmbspec}
The measurement of the CMB frequency spectrum by {\it COBE}/FIRAS places very tight bounds on spectral distortions away from a black-body spectrum \citep{fixsen1996}, limiting possible energy injection into the plasma \citep{wright1994}. 

The effects of energy injection depend crucially on when it occurs, with the time-line roughly divided into three eras. In the earliest era, reactions that change photon number are fast and any injection of photon energy is quickly rethermalized. This leads to only an adjustment of the temperature, hence this is called the ``T era''. Decays during this era do not cause any distortions in the frequency spectrum observed today and are thus allowed. At about $T\approx750\,\rm eV$, double Compton (DC) and Bremsstrahlung (BR) reactions become too slow to significantly change photon number. Injected photons now lead to a spectrum with a chemical potential, giving this the name the ``$\mu$ era''. Finally,  Compton scattering freezes out at $T\approx25\,\rm eV$, making even the redistribution of energy inefficient. Initially we are left with a Comptonized spectrum described by a $y$ parameter, and at even later times photons from decay can be seen directly, contributing to the extra-galactic background. We do not consider $y$ distortion bounds in this work as they rule out only very late decays which are outside of the parameter range of interest.

As shown by \citet{hu1993}, if $\mu(z)\ll1$ for the duration of the energy injection and if Compton scattering is fast compared to DC and BR, the evolution of $\mu$ obeys,
\begin{align}
\label{eq:mu1}
\frac{d\mu}{dt} = \frac{d\mu_s}{dt} - \mu\left(\frac{1}{t_{\rm DC}} + \frac{1}{t_{\rm BR}}\right)
\end{align}
where $1/t_{\rm DC}$ and $1/t_{\rm BR}$ are the scattering rates given by,
\begin{align}
\frac{t_{\rm DC}}{\rm sec} &= \frac{1.9\times10^{35}}{\Omega_bh^2(1-Y_p/2)} \left(\frac{T}{\rm Kelvin}\right)^{-9/2} \\
\frac{t_{\rm BR}}{\rm sec} &= \frac{8.9\times10^{26}}{(\Omega_bh^2)^{3/2}(1-Y_p/2)} \left(\frac{T}{\rm Kelvin}\right)^{-13/4} 
\end{align}
and $d\mu_s/dt$ is a source term given by,
\begin{align}
\label{eq:mu2}
d\mu_s = \frac{3\pi^4\zeta(3)}{2\pi^6-810\zeta(3)^2}\left(3\frac{d\rho_\gamma}{\rho_\gamma}-4\frac{dn_\gamma}{n_\gamma}\right)
\end{align}
which is the chemical potential induced by adding a comoving photon energy density $d\rho_\gamma$ and number $dn_\gamma$. Since ALPs decay to two photons, we have $d\rho_\gamma=-d\rho_\phi$ and $dn_\gamma=-2dn_\phi$. 

In practice, we first evolve Eqn.~\ref{eq:boltz}, which assumes instantaneous thermalization. Then we take the resulting solution for $d\rho_\phi$, $dn_\phi$, and expansion history, and numerically integrate Eqn.~\ref{eq:mu1} to arrive at the actual thermalization history. This procedure is correct to first order in the energy injection. We then compare with the {\it COBE}/FIRAS data which gives a 95\% upper bound of,
\begin{align}
|\mu| < 9\times10^{-5}
\end{align}

The excluded region is shown in orange in Fig.~\ref{fig:mtau_baseline} and is labeled ``CMB Spectral Distortions''. Because Eqn.~\ref{eq:mu2} shows the source term for $\mu$ depends on the fractional energy density injected into the plasma after DC and BR freeze-out, the arguments of Sec.~\ref{sec:scenario} imply that the exclusion region should be one of constant decay time, and indeed it is parallel to $\nu$-dec and BBN start/end lines. The exact division between allowed and disallowed scenarios is close to the transition from the $T$ to the $\mu$ era, as expected because almost any chemical potential is ruled out while almost any injection in the $T$ era is unobservable. These same constraints were calculated by \citet{cadamuro2012} under a further approximation of Eqn.~\ref{eq:mu1} where DC and BR are taken to be infinitely fast until they instantaneously freeze-out at $T\approx750\,\rm eV$. In the ``previous constraints'' panel of Fig.~\ref{fig:mtau_baseline_compare} we reproduce their result, showing that qualitatively this is a very good approximation. The use of Eqn.~\ref{eq:mu1} becomes more important, however, for constraining scenarios with even smaller chemical potentials generated deeper into what is currently called the $T$-era. Two future missions which are predicted to reach such sensitivity are PIXIE and COrE \citep{kogut2011,thecorecollaboration2011}, for which we give forecasts in Sec.~\ref{sec:forecasts}.

\subsubsection{Angular Power Spectrum}
\label{sec:cmbcl}
Measurements of CMB anisotropies have been recently improving, both from the ground \citep{keisler2011a,das2011,story2013a,das2014} and from space \citep{planck2013xvi,planck2013xv}. Better angular resolution and lower noise have tightened up small-scale constraints where the CMB is most sensitive to changes in $N_{\rm eff}$ and $Y_p$, both of which are altered by the decay of ALPs. 

A fully general treatment would include ALPs in the set of Boltzmann equations for calculating the CMB power spectrum, but it turns out this is not necessary for the scales that are currently well measured. For these scales, all of the physical effects of ALPs are in fact identical to changes in $N_{\rm eff}$ and $Y_p$, as long as we assume adiabatic initial conditions. This is essentially because decays must happen early enough, as enforced by the spectral distortion bound discussed in the previous section, which requires the decay happen by $T\approx 750\,\rm eV$ or equivalently $z\approx 3 \times 10^6$. The angular scales well constrained by CMB measurements, roughly $\ell\lesssim 3000$, correspond to physical scales which do not begin to enter the horizon until about $z\approx 3\times 10^5$. At this point two things are different in the ALP scenario as opposed to the standard case 1) the amplitude of neutrino density perturbations upon horizon entry is reduced relative to the photons and 2) the expansion rate is different. However, both are exactly captured in the standard scenario by changing $N_{\rm eff}$. No other scale-dependent changes to spatial perturbations are possible because the relevant scales are still outside of the horizon by the time of the decay. Finally, we note that the altered helium abundance is taken as an input to the CMB spectrum calculation, and in the ALP scenario it is now just at a different value. 

\begin{figure}
\centering
\includegraphics[width=3.3in]{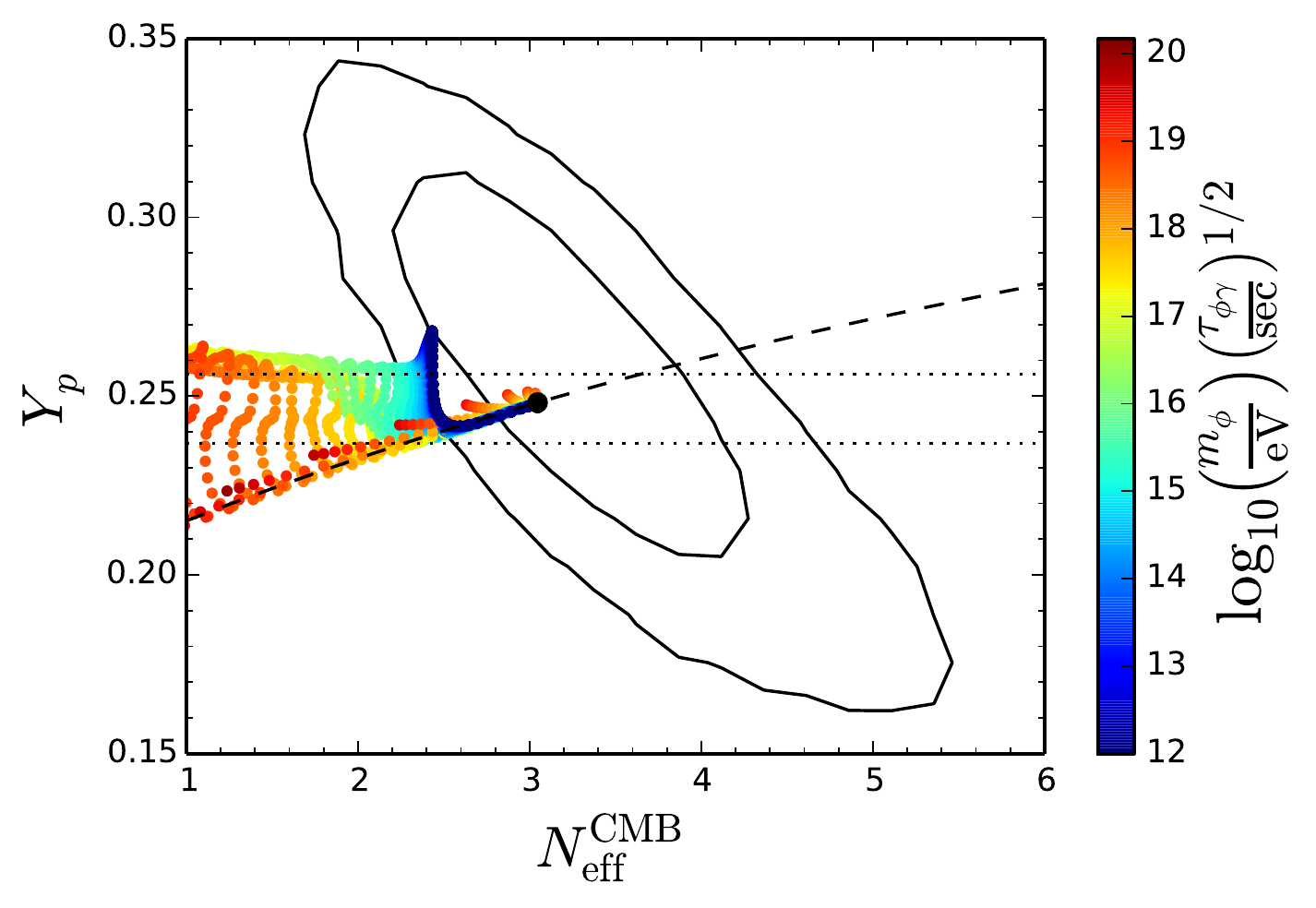}
\caption{The contours show the 1- and 2-$\sigma$ confidence regions for $N_{\rm eff}^{\rm CMB}$ and $Y_p$ from \textit{Planck}+WP+highL. The dotted lines give the 1-$\sigma$ constraint on $Y_p$ from \citet{aver2013}. The dashed line is the relation if standard BBN is assumed, and the dot along this line corresponds to the standard value of $N_{\rm eff}^{\rm CMB}=3.046$. Colored points show values of $N_{\rm eff}^{\rm CMB}$ and $Y_p$ arising from ALP masses and lifetimes sampled from a grid over the entire region shown in Fig.~\ref{fig:mtau_baseline}. They are colored by $m_\phi\sqrt{\tau_{\phi\gamma}}$ which is an important quantity for the CMB constraint since it controls the fractional energy injected into the photons (Eqn.~\ref{eq:mroottau}). The maximum value for in-equilibrium decays of $N_{\rm eff}^{\rm CMB}\approx2.44$ gives the sharp cutoff visible above. Points along the standard BBN consistency line arise from decays happening between neutrino decoupling and the beginning of BBN, and correspond to the island of low helium visible in Fig.~\ref{fig:mtau_series_baseline}.}
\label{fig:neff_yp}
\end{figure}

In their work, \citet{cadamuro2012} took as a CMB constraint a lower bound on $N_{\rm eff}$ as given by WMAP7. The \textit{Planck} data tighten this constraint and are also sensitive to $Y_p$. We use the joint constraint on these two parameters given by the combination of {\it Planck}+WP+highL from \cite{planck2013xvi}, approximating the likelihood as Gaussian and taking just the mean and covariance. In practice, we first calculate $N_{\rm eff}^{\rm CMB}$ and $Y_p$ for a given mass and lifetime (here ignoring uncertainties in $Y_p$ due to $\eta$ and nuclear reaction rate uncertainties which are unimportant at the level of the CMB constraint). We then calculate the $\chi^2$ of these values against the mean and covariance of $N_{\rm eff}^{\rm CMB}$-$Y_p$ from the \textit{Planck} chain. Masses and lifetimes excluded at $>3\sigma$, or equivalently $\chi^2>11.8$, are shown in cyan in Fig.~\ref{fig:mtau_baseline} and labeled ``CMB Anisotropies.'' This region covers a large part of the parameter space corresponding to out-of-equilibrium decays. These decays tend to greatly reduce $N_{\rm eff}^{\rm CMB}$ because ALPs can live long enough to become non-relativistic without becoming Boltzmann suppressed, thus greatly increasing their total energy injection. The majority of the remaining parameter space corresponds to in-equilibrium decay, which is allowed because here there is a lower limit of $N_{\rm eff}^{\rm CMB}\approx2.44$ as discussed in Sec.~\ref{sec:scenario}, compatible with CMB constraints. Fig.~\ref{fig:neff_yp} shows typical values for $N_{\rm eff}^{\rm CMB}$ and $Y_p$ in the ALP scenario as compared to the CMB constraints. 

\subsection{Primordial Abundance Inferences}
\label{sec:bbnconstraints}

The primordial production of the light nuclides during BBN 
occurs from $t\sim 1$ sec to $\sim 3$ minutes, corresponding to $T \sim 1$ MeV to
100 keV \citep[reviewed recently in, e.g.,][]{Iocco2009,Steigman2007,Fields2011}.  
Observations of primordial light element abundances thus probe
new physics at play during this epoch
\citep[e.g., ][]{Pospelov2010}. For our case of an ALP, light element production is affected by
changes to the cosmic expansion rate 
during BBN and to the extrapolation of $\eta$ back from the CMB epoch. For the majority of the parameter space where ALPs decay after neutrino decoupling, they are still present during BBN and hence increase $N_{\rm eff}^{\rm BBN}$ and the expansion rate. Additionally, because the decay decreases $\eta$, fixing $\eta_{\rm CMB}$ to the observed value generally leads to an increased $\eta_{\rm BBN}$. Both effects serve to increase $Y_p$, but partially cancel for D/H and $^7$Li. 
The increase in $N_{\rm eff}^{\rm BBN}$ increases D/H while the increase in $\eta$ has the opposite effect, with the latter about twice as large, leaving an overall reduced D/H. For $^7$Li it is instead the former which wins out. The light-element trends in the
mass-lifetime planes of Fig.~\ref{fig:mtau_series_baseline}
bear out these expectations, as we now see in detail.

In practice we have modified the \texttt{AlterBBN} code of \citet{arbey2012} to include changes to the expansion history and $\eta$ due to ALPs.  Our code assumes the photon spectrum is instantaneously rethermalized, in effect ignoring the possibility that high energy photons from the decay can break apart already-formed nuclei. Bounds due to this phenomenon constitute so called ``photo-erosion'' bounds, discussed in e.g. \cite{cyburt2009} and references therein. We will consider them separately at the end of this section.

Deuterium is observable at $z \sim 3$ in QSO absorption systems, via the $\sim 82$ km/s isotope shift between D and H Lyman absorption lines.  Recent D/H measurements have been reported \cite{cooke2014},
\begin{align}
\label{eq:DHmeasure}
\frac{\rm D}{\rm H} = (2.53 \pm 0.04)\times10^5,
\end{align}
which represents a factor $\sim 3$ improvement in precision.  These bounds are now so tight as to place the measurement errors on level footing with uncertainties associated with nuclear reaction rates and with a determination of $\eta$ from the CMB. To account for these uncertainties, we first consider the joint likelihood for the CMB, D/H, and nuclear reaction rate measurements, which can be written as
\begin{align}
-\log\mathcal{L} = &\frac{\left[{\rm DH}(m_\phi, \tau_{\phi\gamma}, \eta, \alpha_i)-\overline{\rm DH}\,\right]^2}{2\sigma^2_{\rm MEAS}}\nonumber\\
&+\log\mathcal{L_{\rm CMB}}(\eta,N_{\rm eff},\Omega)\nonumber\\
&+\log\mathcal{L_{\rm NUCL}}(\alpha_i, \Omega')
\end{align}
where $\overline{\rm DH}\pm\sigma_{\rm MEAS}=(2.53\pm0.04)\times10^5$ as per Eqn.~\ref{eq:DHmeasure}, $\eta$ and $N_{\rm eff}$ are evaluated at the CMB epoch but we omit the label for brevity, $N_{\rm eff}=N_{\rm eff}(m_\phi, \tau_{\phi\gamma})$ is uniquely set by the mass and lifetime, $\alpha_i$ are parameters describing the nuclear reaction rates, and $\Omega$ and $\Omega'$ are any remaining cosmological and nuisance parameters. $N_{\rm eff}$ is important and appears explicitly because it is both dependent on the ALP parameters and its measurement from the CMB is significantly degenerate with $\eta$. We next analytically marginalize over all parameters other than $m_\phi$ and $\tau_{\phi\gamma}$ under the assumption that these other parameters have Gaussian posterior likelihoods and that D/H depends linearly on $\eta$. This gives 
\begin{align}
\label{DHlike}
-\log\mathcal{L} = &\frac{\left[{\rm DH}(m_\phi, \tau_{\phi\gamma}, \bar\eta + r \sigma_\eta \frac{N_{\rm eff}-\overline{N_{\rm eff}}}{\sigma_{N_{\rm eff}}}, \overline \alpha_i)-\overline{\rm DH}\,\right]^2}{2\left[\sigma^2_{\rm MEAS}+\sigma^2_{\rm NUCL}+\sigma^2_{\rm ETA}   \right]}
\end{align}
with 
\begin{align}
\sigma^2_{\rm ETA} = \left(\frac{d{\rm DH}}{d\eta}\sigma_\eta\right)^2(1-r^2)
\end{align}
where $\bar \eta \pm \sigma_\eta$ and $\overline{N_{\rm eff}} \pm \sigma_{N_{\rm eff}}$ are the mean and standard deviation of the posterior likelihoods from the CMB with all other parameters marginalized over, and $r$ is the correlation coefficient between $\eta$ and $N_{\rm eff}$. The presence of $r$ in this equation can be understood by considering the $r=1$ case, which would imply that CMB measurements could turn a fixed $N_{\rm eff}$ into a perfect determination of $\eta$; the quantity above at which the D/H prediction is evaluated, $\bar\eta + r \sigma_\eta (N_{\rm eff}-\overline{N_{\rm eff}})/\sigma_{N_{\rm eff}}$, is the mean of this determination. Because in our case $N_{\rm eff}$ is fixed by the mass and lifetime, it would mean $\eta$ is also fixed, leading to no extra uncertainty in D/H. In reality, we find $r\approx0.4$ from the \textit{Planck} measurements. We make one further approximation which is that neither the D/H derivative nor the nuclear reaction rate uncertainty depends on the values of mass and lifetime or the fact that $\eta$ evolves with time in the ALP scenario, which we have checked is sufficient. We find $\sigma_{\rm NUCL}=4.5\times10^{-7}$ using \texttt{AlterBBN} which takes the $\alpha_i$ to be principal components in the nuclear reaction rate parameter space \citep{fiorentini1998}. Numerically evaluating the D/H derivative and taking posterior likelihoods from \textit{Planck}+WP+highL, we find $\sigma_{\rm ETA}=6.9\times10^{-7}$. When  added in quadrature these lead to an effective deuterium constraint of 
\begin{align}
\frac{\rm D}{\rm H} = (2.53 \pm 0.091)\times10^5
\end{align}
which is meant to be compared to a theoretical prediction calculated for the particular values of $\eta$ and $\alpha_i$ given in Eqn.~\ref{DHlike}.

The effects of these D/H constraints on the
$(m_\phi,\tau_{\phi \gamma})$ plane appear in 
Fig.~\ref{fig:mtau_series_baseline}.
We see that the effect of an ALP is always to decrease
D/H due to the ALP's
effective increase of $\eta_{\rm BBN}$ winning out over the increase in $N_{\rm eff}^{\rm BBN}$. Moreover, we see that
the high precision of the D/H measurements
leads to a tight constraint on the ALP space
in all regions where the decays occur after 
neutrino decoupling.  Indeed, D/H is now a very
powerful probe of ALPs.


The primordial $^4\rm He$ abundance is inferred astronomically from
observations of emission spectra of highly ionized gas
in primitive nearby dwarf galaxies,
i.e., in low-metallicity
extragalactic HII regions.
The primordial abundance is traditionally inferred by
extrapolation to zero metallicity.
To derive helium and metal abundances from the observed spectra 
requires characterization of the thermodynamic
properties of the emitting gas (i.e., temperature, density).
The analysis of 
\citep{aver2013} 
derives these quantities simultaneously in a self-consistent
manner, and finds a primordial abundance
\begin{align}
Y_p = 0.2465 \pm 0.0097
\end{align}
where the uncertainty is quantified with an MCMC analysis.
We adopt this as our fiducial primordial $^4\rm He$ constraint.
Using a similar data set,
\citet{izotov2010a} give a helium constraint of, 
$Y_p = 0.254 \pm 0.003$
where the errors are derived in a less conservative manner.
In Fig.~\ref{fig:mtau_series_baseline} we see that, 
as expected, the
effect of an ALP is to increase $Y_p$ for almost all of
the parameter space where the decays occur after neutrinos decouple.
However, the constraints are not as strong
as those of D/H.
There is an island of parameter space
around $m_\phi > 100$ MeV
and $\tau_{\phi \gamma} \sim 1$ sec
where $Y_p$ decreases.
This region corresponds to decays happening between neutrino decoupling and the start of BBN. Here we have $\eta_{\rm BBN}=\eta_{\rm CMB}$ and $N_{\rm eff}<3$, the latter of which serves to decrease $Y_p$. It is interesting to note this low helium region does not extend along the entire constant decay-time contour, cutting off once we enter the in-equilibrium decay side. This occurs because in-equilibrium decays reduce the ALP energy density more slowly than do out-of-equilibrium ones, and thus the entire decay cannot fit in the short time between neutrino decoupling and BBN. 

Finally, the primordial $^7\rm Li$ abundance is inferred from
observations of the atmospheres of low-metallicity (extreme Population II)
stars
in the Galactic stellar halo.
Down to a metallicity of $({\rm Fe/H}) \sim 10^{-2.8}({\rm Fe/H})_\odot$,
these stars have lithium abundances that are the same
to within a small scatter
consistent with observational errors.  The independence of 
Li with Fe in this ``Spite plateau'' indicates that lithium is primordial
\citep{Spite1982}, and implies a primordial abundance
\begin{align}
\label{eq:Li_p}
\frac{\rm Li}{\rm H} = (1.6 \pm 0.3) \times 10^{-10}
\end{align}
\citep{Sbordone2010}.
At lower metallicity, however, the Li/H abundance
scatter increases dramatically, but always below
the Spite plateau value.
This suggests that in these very metal-poor stars some
lithium destruction has occurred; the reason for this remains unclear.

The astronomically-inferred lithium abundance in Eqn.~\ref{eq:Li_p}
is inconsistent with the primordial value expected
from standard BBN theory combined with CMB determinations of $\eta$.
The observed Li/H value is low at the $\sim 5 \sigma$ level.
This is the ``lithium problem'' \citep[reviewed in, e.g.,][]{Fields2011}.
Stellar astrophysics uncertainties may be the origin of the problem,
but solutions to date require fine tuning and do not explain the
observed Li/H ``meltdown'' at very low metallicities.
A more radical and intriguing solution is the presence of new physics
during or after BBN.  The challenge for such scenarios is to 
reduce lithium substantially without drawing other light elements--particularly
deuterium--from their concordant primordial abundances. The ALP scenario tends to aggravate the problem by increasing lithium slightly, as
seen in Fig.~\ref{fig:mtau_series_baseline}, and as expected due to
the ALP effect on $\eta$.
Thus, awaiting a resolution to the lithium problem, we do not consider lithium bounds. 

In closing we note that our calculations have neglected the effects
of photoerosion of the light elements.
This occurs 1) when the ALP mass exceeds light-element
binding energies $m_\phi \ga B \sim 10$ MeV, and 2) for decay time scales
long enough so that the decay photons interact with light elements
before thermalization.  This leads to some deuterium destruction via
$\gamma d \rightarrow n p$, but a net production due to e.g., $\gamma ^4{\rm He} \rightarrow d d$. Thus constraints arise from D/H, $Y_p$, and $^{3}{\rm He}$/D \citep{ell1984,kaw1995,cyb2003}.  These were recently computed for purely electromagnetic
decays by \citet{cyburt2009}, assuming that the decay photons
provide a negligible contribution to the energy density and thus expansion rate.
In this case, the constraints are only important for $\tau_X \ga 10^4$ sec
and $m_X \ga 10^7$ eV, with $X$ the decaying particle.
This regime shows a $Y_p$ drop due to photoerosion
and a corresponding D/H increase. These trends could potentially bring $Y_p$ and D/H predictions back into agreement with observations, but would require a more detailed calculation. Since the regions of parameter where this can happen are already ruled out by CMB observations, we ignore the effects of photo-erosion.

\subsection{Laboratory}
\label{sec:lab}
Laboratory bounds on ALPs come from a variety of different experimental setups. At lower masses, roughly $m \lesssim\rm eV$, some examples include photo-regeneration experiments (``shining light through walls''), microwave cavities, and helioscopes \citep[for a review, see][]{hewett2012,essig2013,olive2014}. For the larger masses considered here, the best constraints come from electron-positron colliders and beam dumps. 

The presence of the ALP-photon interaction allows for the possibility of single-photon final states at electron-positron colliders. Early interpretation of searches for these events in terms of constraints on the ALP coupling $g_{\phi\gamma}$ was done by \citet{masso1995}. Both \citet{kleban2005} and \citet{mimasu2014} have further shown the ability of current and future colliders to improve these bounds. Here we reproduce the constraint from LEP given by \citet{mimasu2014} of,
\begin{align}
g_{\phi\gamma} < 4.5\times10^{-4} \,\rm GeV^{-1},
\end{align}
valid in the entire lifetime range considered here. The excluded region is labeled ``Collider'' and shown in magenta in Fig.~\ref{fig:mtau_baseline}. 

Additional constraints come from beam dump experiments, where ALPs would be produced in the beam dump, penetrate through shielding, then decay to photons which can be detected by a downstream detector. We use the constraints from \citet{bjorken1988} which find that at 95\% confidence,
\begin{align}
m_\phi \tau_{\phi\gamma} > 1.4 \, \rm keV \, sec. 
\end{align}
This is labeled ``Beam Dump'' and shown in red in Fig.~\ref{fig:mtau_baseline}. 

\subsection{Globular Clusters and SN1987A}
\label{sec:astrophysical}
ALPs offer a new means for energy loss from stars if they can both be produced in stellar interiors and have sufficiently weak interaction strengths to subsequently escape. In the case of SN1987A, the energy loss can affect the duration of the neutrino pulse from the handful of neutrinos which were detected, placing constraints on the ALP interaction strength \citep{masso1995,masso1997}. These bounds are reproduced in Fig.~\ref{fig:mtau_baseline}. We note that they assume the ALP has only a two-photon coupling, although constraints based on other couplings exist. Energy loss can also affect the duration of the red giant phase and of the horizontal branch, leading to a different observed ratio of such stars in globular clusters. We use the bounds from \citet{cadamuro2012} based on arguments of \citet{raffelt1988} and \citet{raffelt1996}. These are also reproduced in Fig.~\ref{fig:mtau_baseline}.

\section{Discussion}
\label{sec:discussion}
\subsection{The MeV-ALP Window}
\label{sec:axion}

An interesting feature of the exclusion regions prior to this work is the allowed window bounded on all sides near $m\sim1\,\rm MeV$ and $\tau\sim 100\,\rm ms$ corresponding to an ALP decay during BBN. We will henceforth call this the MeV-ALP window. It can be seen in the left panel of Fig.~\ref{fig:mtau_baseline_compare} as well as in \citet{hewett2012,mimasu2014}. Further interest is driven by the fact that a particle in this window could actually be a DFSZ axion, in which case its symmetry breaking scale is close to the electroweak scale. One of the main conclusions of this work is to show that this region is now, in fact, ruled out by the combination of CMB+D/H measurements.

We first briefly review two relevant generic axion models, referred to as the KSVZ and DFSZ models. Both models introduce a new global $U(1)$ symmetry which is approximately broken at some energy scale $f_\phi$ giving rise to an axion with mass $m_\phi$. The symmetry breaking scale is related to the axion mass by non-perturbative effects and given by 
\begin{align}
m_\phi = \frac{\sqrt z}{1+z} \frac{m_\pi f_\pi}{f_\phi}
\end{align}
where $m_\pi$ is the pion mass, $f_\pi$ its decay constant, $z=m_u/m_d$ the ratio of up to down quark masses. Axion models differ in what other new fields are introduced to implement the symmetry breaking and how these, as well as standard model fields, transform under the new $U(1)$. The KSVZ model \citep{kim1979,shifman1980} has the standard model fermions neutral, whereas in the DFSZ model \citep{zhitnitskij1980,dine1981} they can carry $U(1)$ charge. These model dependent choices in turn affect the axion's effective photon coupling which arises from fermion loops, ultimately leading to a consistency relation between axion mass and photon lifetime which can be written as,
\begin{align}
\label{eq:axionlife}
\tau_{\phi\gamma} &= \left[ \left(\frac{E}{N}-\frac{2(4+z)}{3(1+z)}\right)^{-1} \frac{\sqrt{z}}{1+z} \frac{16\pi^{3/2}}{\alpha} f_\pi m_\pi \right]^{2} m_\phi^{-5}
\end{align}
with the model dependence captured by the $E/N$ factor. In evaluating this relation, we will adopt fixed values of $m_\pi=135\,\rm MeV$, $f_\pi=92\,\rm MeV$, and $z=0.56$, ignoring small uncertainties that lead to roughly a 10\% uncertainty in the axion mass \citep{cadamuro2011a,beringer2012}.

The KSVZ model has $E/N=0$, so that we have, 
\begin{align}
 \left(\frac{\tau_{\phi\gamma}}{\rm sec}\right) &= 6.20\times10^{4} \left(\frac{m_\phi}{\rm eV}\right)^{-5}
\end{align}
The DFSZ model we consider here has $E/N=2$, hence we refer to it as the ``DFSZ-EN2'' model, with consistency relation,
\begin{align}
\label{eqn:dfszen2}
\left(\frac{\tau_{\phi\gamma}}{\rm sec}\right) &= 2.66\times10^{5} \left(\frac{m_\phi}{\rm eV}\right)^{-5}
\end{align}
These two consistency relations are shown as the dashed lines in Fig.~\ref{fig:mtau_baseline}. Other values for $E/N$ are possible, but the DFSZ-EN2 has the distinction of having a particularly weak coupling because $E/N\approx2(4+z)/3(1+z)$ and so these terms nearly cancel in Eqn.~\ref{eq:axionlife} \citep{kaplan1985,cheng1995}. This weaker coupling means that the DFSZ-EN2 is consistent with the collider bounds over a larger range of masses as compared to the KSVZ. Ultimately it is that the consistency relation passes through the MeV-ALP window which motivates our interest in this model. The lower mass limit for the DFSZ-EN2 in the MeV-ALP window is around $m_\phi\sim 200\,$keV, corresponding to $f_\phi \sim 30\,$GeV, less than an order of magnitude from the electroweak scale $v_{\rm weak}\sim 246 \, \rm GeV$ where the axion was initially thought to lie. 


While it is interesting that this mass range for the DFSZ-EN2 was previously allowed, this part of parameter space for the DFSZ-EN2, and more generally the entire MeV-ALP window, is now ruled out by the combination of CMB+D/H data. This region corresponds to in-equilibrium decays hence it gives $N_{\rm eff}^{\rm CMB}=2.44$. The decay happens essentially in the middle of BBN, increasing $N_{\rm eff}^{\rm BBN}$ and $\eta_{\rm BBN}$, which, as discussed previously, increases $Y_p$ and decreases D/H. The decrease in $N_{\rm eff}^{\rm CMB}$ and increase in $Y_p$ moves along the degeneracy direction for CMB measurements, and is allowed even by our updated CMB constraints coming from \textit{Planck} (see Fig.~\ref{fig:neff_yp}). It is in combination with the D/H constraints that the MeV-ALP window is closed, with the best fitting model within the window ruled out at about $3.5\sigma$. If we replace the CMB constraint from \textit{Planck} with previous measurements from the combination of WMAP, ACT, and SPT, the window is ruled out at a similar significance. This is despite the 20\% tighter $\eta$ constraint from \textit{Planck} because the central value also shifts lower, increasing D/H back towards the measured value. Conversely, replacing the D/H measurement with previous bounds \textit{does} open the MeV-ALP window again, as seen in the left panel of Fig.~\ref{fig:mtau_baseline_compare}. It is thus the new bounds from \citet{cooke2014} that are the key improvement.

\begin{figure}
\centering
\includegraphics[width=3in]{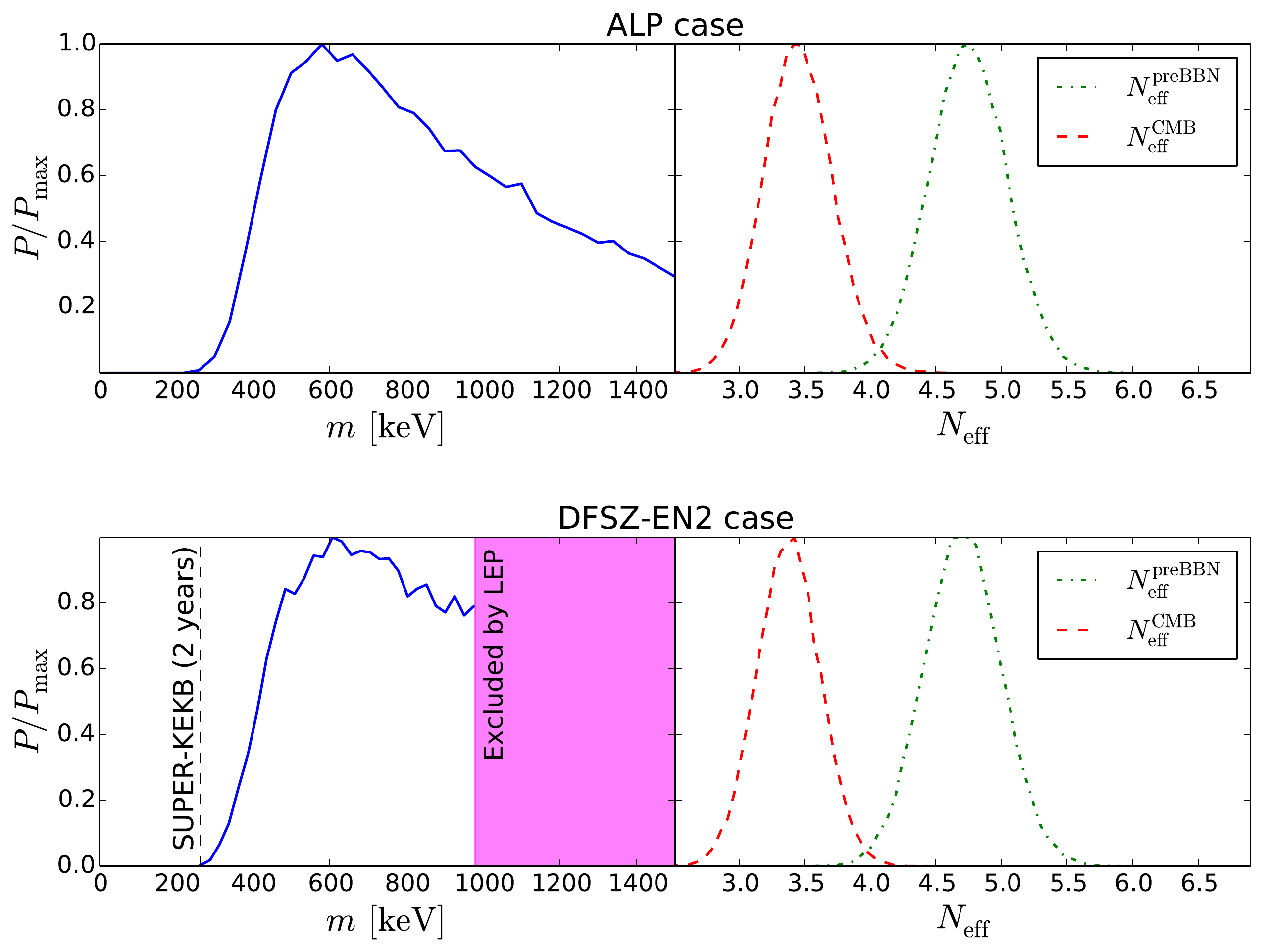}
\caption{Parameter constraints in the MeV-ALP region of parameter space when we also allow extra radiation present besides neutrinos and the ALP. Our likelihood includes all of the constraints shown in Fig.~\ref{fig:mtau_baseline}. In the ALP case (top panel) the lifetime is marginalized over whereas in the DFSZ-EN2 case (bottom panel) it is fixed by the consistency relation (Eqn.~\ref{eqn:dfszen2}). The vertical dashed line is a forecast for SUPER-KEKB, showing that it could close the remaining allowed parameter window or detect a particle there. We show $N_{\rm eff}$ evaluated both prior to BBN when neutrinos, extra radiation, and the $\sim$MeV ALP (which here adds $4/7$ to $N_{\rm eff}$) contribute, and at the CMB epoch after the ALP has decayed.}
\label{fig:mass}
\end{figure}

\subsection{A Loophole in the Presence of Extra Radiation}
\label{sec:loophole}

\begin{figure}
\centering
\includegraphics[width=3.2in]{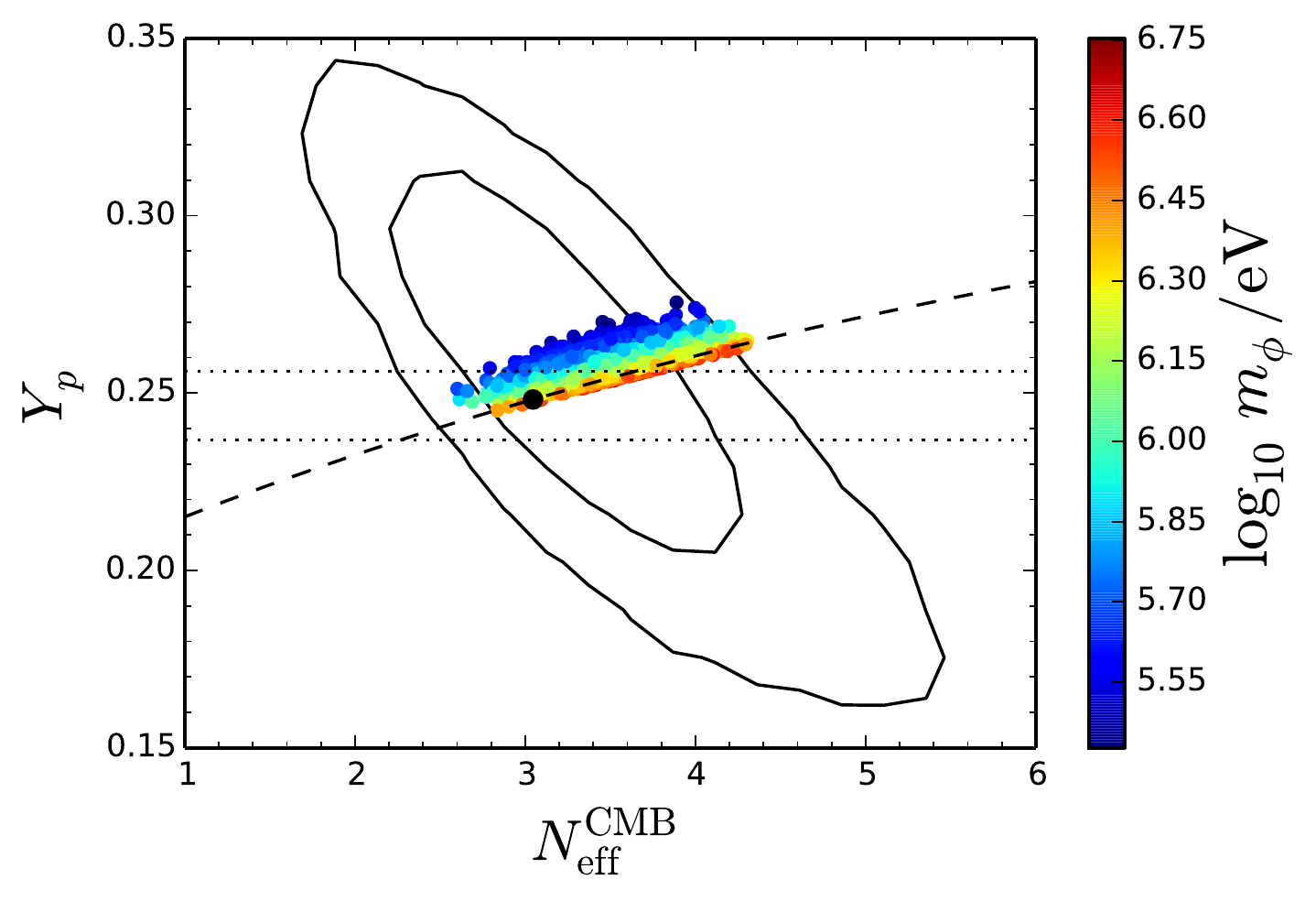}
\includegraphics[width=3.2in]{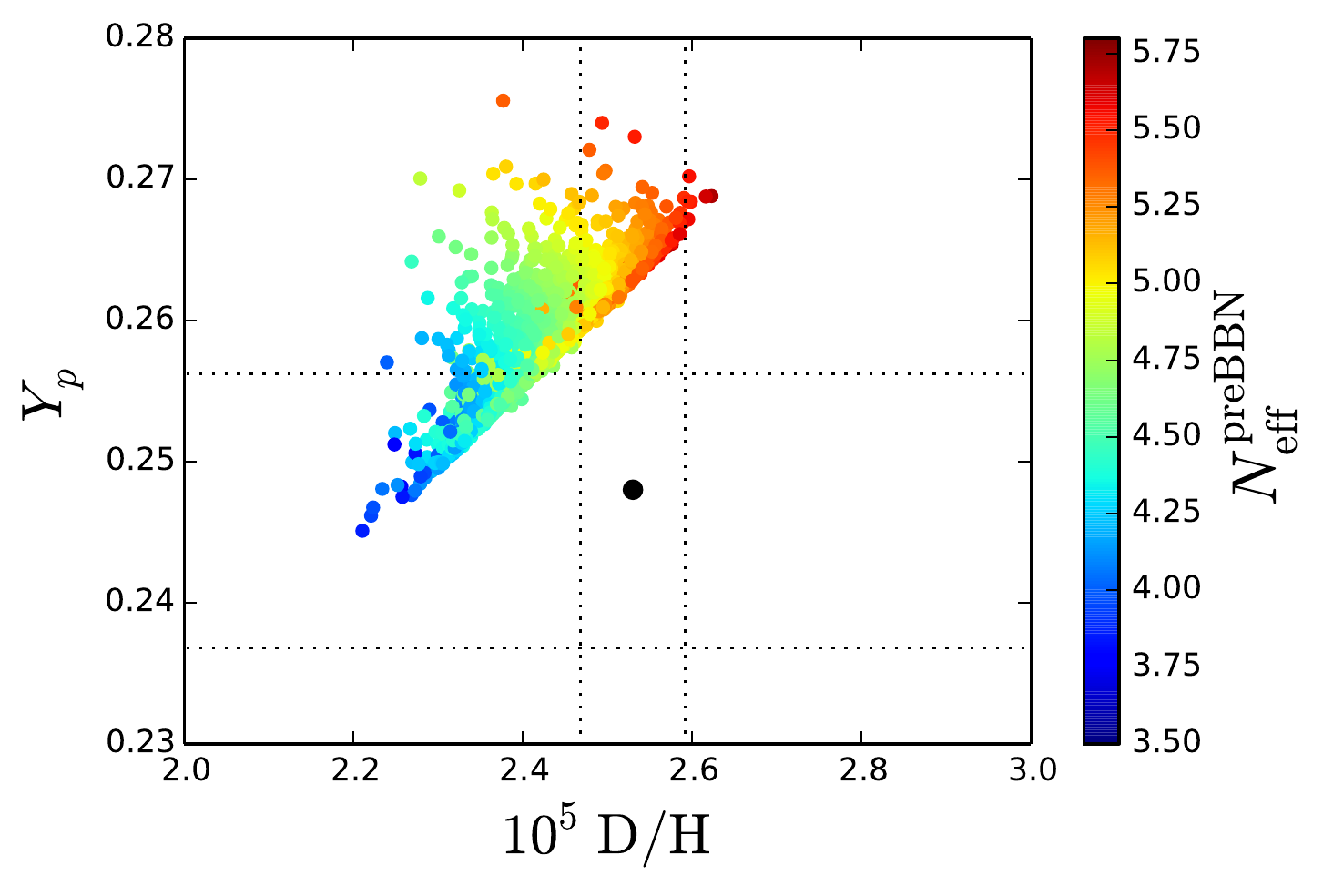}
\caption{(\textbf{Top}) The contours show the 1- and 2-$\sigma$ confidence regions for $N_{\rm eff}^{\rm CMB}$ and $Y_p$ from \textit{Planck}+WP+highL. The dotted lines give the 1-$\sigma$ constraint on $Y_p$ from \citet{aver2013}. The dashed line is the relation if standard BBN is assumed, and the dot along this line corresponds to the standard value of $N_{\rm eff}^{\rm CMB}=3.046$. Colored points show values of $N_{\rm eff}^{\rm CMB}$ and $Y_p$ arising from ALP masses and lifetimes taken from the $\rm \Lambda$CDM+$\Delta N_{\rm eff}$+ALP chain described in Sec.~\ref{sec:loophole}. They are colored by $m_\phi$ which controls decay time and can alter BBN but otherwise does not affect $N_{\rm eff}^{\rm CMB}$. (\textbf{Bottom}) Same as the top panel, but with $\rm 10^5 D/H$ shown on the x-axis. The vertical dotted lines give the 1-$\sigma$ constraint from \citet{cooke2014}. Points are colored instead by $N_{\rm eff}^{\rm preBBN}$. Sufficiently tight constraints around the standard value (black dot) could rule out the ALP scenario even in the presence of extra radiation.}
\label{fig:loophole}
\end{figure}

While the MeV-ALP region is now excluded by the CMB+D/H measurements, this result depends on the assumption of having no extra radiation besides neutrinos and the ALP. In some scenarios, for example as predicted by the string axiverse \citep{arvanitaki2009}, it is natural to have many ALPs, some of which could also contribute to $N_{\rm eff}$ but be light enough to remain otherwise invisible. Motivated by this possibility, we explore constraints when in addition to the ALP mass and lifetime, we also allow an extra arbitrary addition to $N_{\rm eff}$. 

The MeV-ALP region is ruled out largely because it predicts too low an abundance of primordial deuterium. An addition to $N_{\rm eff}^{\rm BBN}$ increases D/H and can bring it back into agreement with measurements. The penalty is a further increase in $Y_p$, but because the helium constraints are not as tight as D/H, an allowed window now opens up again. 

We explore this window with Markov Chain Monte Carlo (MCMC) \footnote{\url{https://github.com/marius311/cosmoslik}}. We run two MCMC chains, one for the ALP case where both mass and lifetime are free parameters, and another for the DFSZ-EN2 case with only the mass free and the lifetime given by Eqn.~\ref{eqn:dfszen2}. In both cases we also leave free the quantity we call $\Delta N_{\rm eff}$ which controls any extra relativistic energy density at some early time before BBN, meaning that $N_{\rm eff}^{\rm preBBN}=3+4/7+\Delta N_{\rm eff}$. The ALP contributes 4/7 because it is one bosonic degree of freedom and is fully thermalized in all regions of parameter space explored by the chain. The likelihood includes all of the bounds in Fig.~\ref{fig:mtau_baseline}. In either the ALP of DFSZ-EN2 cases, we find that the MeV-ALP window is again allowed and the best-fitting model consistent with all of the data at $\lesssim 1\sigma$. 

The mass posterior distributions for the ALP and DFSZ-EN2 chains are shown in Fig.~\ref{fig:mass}. In both cases masses below 200\,keV are excluded. Masses above 1\,MeV in DFSZ-EN2 case are excluded by the collider bound, but are allowed in the ALP case because models can evade this constraint by having a smaller photon coupling. We also show a forecast for a next generation electron-positron collider SUPER-KEKB after two years of integration (discussed in Sec.~\ref{sec:forecasts}) which can probe down to almost exactly the 200\,keV minimum. 


The corresponding likelihoods for $N_{\rm eff}$ are given in the right panel. These show that the data accommodate the ALP scenario by initially having $N_{\rm eff}^{\rm preBBN}\sim4.7$ and diluting this down to $N_{\rm eff}^{\rm CMB}\sim3.4$ via the ALP decay. Constraints on the extra radiation are $\Delta N_{\rm eff}=1.13\pm0.30$. Thus, one or (marginally) two extra neutrino-like particles allow for an ALP in the MeV-ALP window. Alternatively, the decay of such an ALP can hide the existence of one or two additional neutrino-like particles from the tight CMB constraints which would otherwise rule them out. A similar loophole allowing for extra radiation has been proposed by \citet{ho2013}.

We next test the extent to which the data prefer these extended models. We perform a simple test using best-fit $\chi^2$ values given in Tab.~\ref{tab:chi2}. If the $\chi^2$ for the extended model decreases significantly as compared to $\rm \Lambda CDM$, then roughly that model is preferred. Although all bounds from Fig.~\ref{fig:mtau_baseline} are included in the fit, we only give $\chi^2$ for those which are not hard cutoffs. When using the combination of \textit{Planck}+D/H+$Y_p$, we find the baseline data choice slightly disfavors both the DFSZ-EN2 and ALP models. The only case where there is a preference for the extend model is in the ALP case when using the helium constraint from \citet{izotov2010a}. Here we find an improvement in $\chi^2$ of 5.81 when we have added 3 new free parameters, something we expect to happen by chance only 12\% of the time. If the high helium value inferred by \citet{izotov2010a} is confirmed, then this scenario is a natural explanation as it can increase helium compared to the standard value while keeping the deuterium abundance and $N_{\rm eff}^{\rm CMB}$ roughly unchanged.

\begin{table*}
\centering
\caption{Best-fit parameters}
\begin{tabular}{l|ccccccc}
\hline\hline
& $m\,[\rm keV]$ & $\tau\,[\rm ms]$ & $N_{\rm eff}^{\rm preBBN}$ & $N_{\rm eff}^{\rm CMB}$ & $10^5$ D/H & $Y_p$ & $10^{10} \, \rm Li^7$ \\
\hline
$\Lambda$CDM &  &  & 3 & 3.046 & 2.56 & 0.247 & 4.58 \\
$\Lambda$CDM+$\Delta N_{\rm eff}$+ALP & 1936 & 4.6 & 4.73 & 3.52 & 2.46 & 0.255 & 5.02 \\
$\Lambda$CDM+$\Delta N_{\rm eff}$+DFSZ-EN2 & 734 & 6.2 & 4.61 & 3.30 & 2.45 & 0.258 & 5.15
\end{tabular}
\label{tab:bf}
\end{table*}

\begin{table*}
\centering
\caption{Best-fit $\chi^2$}
\begin{tabular}{l|cccccc}
\hline\hline
& Planck(2) & Cooke(1) & Aver(1) & Izotov(1) & Planck+Cooke+Aver(4) & Planck+Cooke+Izotov(4)  \\
\hline
$\Lambda$CDM & 0.96 & 0.10 & 0.00 & 5.62 & 1.06 & 6.68\\
$\Lambda$CDM+$\Delta N_{\rm eff}$+ALP & 0.22 & 0.57 & 0.74 & 0.08 & 1.53 & 0.87\\
$\Lambda$CDM+$\Delta N_{\rm eff}$+DFSZ-EN2 & 0.02 & 0.91 & 1.43 & 1.86 & 2.35 & 2.79
\end{tabular}
\label{tab:chi2}
\end{table*}

\subsection{A Simple Expression for Exclusion Bounds}
Given the improved constraints from CMB+D/H measurements, we suggest a simple expression for ALP bounds which can be adopted by those who prefer a simpler picture than the many probes shown in Fig.~\ref{fig:mtau_baseline}. The CMB+D/H data alone now essentially rule out any energy injection after neutrino decoupling, giving allowed parameters of
\begin{align}
\frac{m_\phi}{\rm eV}>10^7 \; {\rm and}\; \frac{\tau_{\phi\gamma}}{\rm sec}<10^{-2}.
\end{align}
This assumes no extra radiation besides ALPs, and is valid roughly until masses become small enough or lifetimes long enough that decays happen after CMB last scattering. These late decays are analyzed in more detail by \citet{cadamuro2012}, who find approximately
\begin{align}
\frac{m_\phi}{\rm eV}<10^1 \; {\rm or}\; \frac{\tau_{\phi\gamma}}{\rm sec}>10^{24},
\end{align}
are once again allowed.

\subsection{Forecasts}
\label{sec:forecasts}
Measurements relevant for placing bounds on ALP parameters have been recently improving and will continue to do so in the near future. It is expected that several probes will soon have the sensitivity to further test the MeV-ALP window. In this section we compute forecasts for some of them. 

Currently CMB anisotropies alone are not enough to rule out the MeV-ALP window where there is a maximum of $N_{\rm eff}=2.44$ and an increase in $Y_p$, but which lie along the CMB degeneracy direction and are thus allowed.  \citet{abazajian2013} show that a Stage-IV CMB experiment could measure $N_{\rm eff}^{\rm CMB}$ to within 0.02 at 1-$\sigma$. Given such tight constraints, the arguments of Sec.~\ref{sec:cmbcl} may need to be revisited; while it is true that the ALP decays before any modes relevant for the CMB enter the horizon, the difference is only an order of magnitude in scale factor. Assuming any such corrections do not provide loopholes, if a Stage-IV CMB measurement found a value of $N_{\rm eff}^{\rm eff}$ consistent with the standard value of 3.046, the MeV-ALP window would be strongly ruled out \footnote{Their forecast assumes standard BBN, however we believe it unlikely that freeing $Y_p$ could degrade it so much as to allow 2.44}. For the CMB constraints, however, there will always be the possibility of extra radiation exactly canceling the dilution due to the ALP decay (as in Sec.~\ref{sec:loophole}). 

Helium and deuterium measurements will continue to improve as more systems are discovered and systematic errors are better understood. Additionally, D/H measurements can be significantly improved by better measuring nuclear reaction rates in the laboratory. Fig.~\ref{fig:loophole} shows that in the D/H-$Y_p$ plane, the $\Lambda$CDM+$\Delta N_{\rm eff}$+ALP scenario is not continuous with the standard model. This allows for sufficiently tight $Y_p$ and D/H constraints around the standard values to rule out the presence of ALPs, independent of assumptions about extra radiation. We find that the minimum requirement for all points in the $\Lambda$CDM+$\Delta N_{\rm eff}$+ALP chain to be ruled out at $>3\sigma$ by the combination of $Y_p$ and D/H measurements is a factor of two improvement in the D/H error and a factor of three improvement in the $Y_p$ error bar. We note the former is possible by eliminating the uncertainty due to nuclear reaction rates alone.

On the laboratory side, in Sec.~\ref{sec:lab} we used the constraint from LEP which limited $g_{\phi\gamma}\lesssim 4.5 \times 10^{-4} \,\rm GeV^{-1}$. If a search for single-photon events were performed using the entire $1000 \,{\rm fb}^{-1}$ of currently existing KEKB data and the standard model background was found, forecasts from \citet{kleban2005} show that the constraints could improve to $g_{\phi\gamma}\lesssim 10^{-6} \,\rm GeV^{-1}$. Similar improvement could come from reinterpreting the constraints on dark photons from $500 \, {\rm fb}^{-1}$ of BABAR data given in \cite{thebabarcollaboration2014} in terms of ALPs. Attempting either of these is outside of the scope of this paper, but could make significant improvements in the mass bounds shown in Fig.~\ref{fig:mass}. SUPER-KEKB, an ongoing upgrade to KEKB, plans to improve on the integrated luminosity of KEKB by a factor of ten with two years of integration, and by a factor of fifty with ten years \citep{abe2010}. Taking the constraint on $g_{\phi\gamma}$ to scale with the square root of the integrated luminosity, we find that within the first two years, SUPER-KEKB can rule out the last remaining part of the MeV-ALP window through which the DFSZ-EN2 passes (the mass limit forecast is shown in Fig.~\ref{fig:mass}). The full ten year forecast is exactly enough to close the MeV-ALP window entirely, bringing the collider bound up to $g_{\phi\gamma}\lesssim 10^{-6} \,\rm GeV^{-1}$ where the SN1987a constraint begins. This simple forecast is in broad agreement with a more sophisticated calculation given by \citet{mimasu2014}. 

Finally, PIXIE is a proposed mission which would greatly improve constraints on CMB spectral distortions \citep{kogut2011}. Expected bounds on the $\mu$ parameter are,
\begin{align}
|\mu|<5\times10^{-8}.
\end{align}
Bounds due to spectral distortion constitute constant decay-time boundaries (Sec.~\ref{sec:scenario}), and we find that the PIXIE forecast moves up the exclusion region by only a factor of five in decay time as compared to FIRAS. This is much less than needed to reach the MeV-ALP window. Evidently it is very difficult to constrain decays happening very much into the $T$ era using spectral distortions. 

\section{Conclusion}
We have shown how cosmological, astrophysical, and laboratory bounds can provide complementary constraints in the mass-lifetime parameter space of axions and ALPs. We have updated the work of \citet{cadamuro2012} with constraints from the \textit{Planck} satellite and the latest inferences of primordial D/H and helium abundances, and provided a more detailed calculation of spectral distortions. The most important change is that CMB+D/H constraints now rule out the entire region corresponding to decays happening after neutrino decoupling but before CMB last scattering. This includes closing the MeV-ALP region of parameter space which we have also shown can correspond to a type of DFSZ axion. The presence of additional radiation can relax the exclusion regions and once again allow the MeV-ALP window. Although it is allowed in this case, including such a particle slightly degrades the overall fit to the CMB+BBN data if our most robust data combination is used. Alternatively, this model can provide a natural explanation for a high value of helium such as found by \citet{izotov2010a}. Forecasts for future primordial abundance measurements and for SUPER-KEKB are promising; both have the ability to test the MeV-ALP window even in the presence of extra radiation. A detection by either would be very exciting. Even the null result, however, would signify a new level of precision in our understanding of the contents of the primordial plasma.

\begin{acknowledgments}
We thank Kevin Cleary, Francis-Yan Cyr-Racine, Rouen Essign, Brent Follin, Nemanja Kaloper, Yury Kolomensky, Markus Luty, Javier Redondo, and McCullen Sandora for helpful conversations and input.  We acknowledge support from NSF awards No. 1213801.

\end{acknowledgments}

\bibliography{DMDecay,DMDecay2}

\end{document}